\documentclass[a4paper,12pt]{article}
\usepackage[utf8]{inputenc}
\usepackage{cancel}
\usepackage{ulem}
\usepackage{amsfonts}
\usepackage{amssymb}
\usepackage{graphicx}
\usepackage{amsmath}
\usepackage{enumerate}
\usepackage{mathtools}
\usepackage{subfig}
\usepackage{color}
\usepackage{float}
\usepackage{here}
\usepackage{cite}
\usepackage{mathrsfs}
\usepackage{float,epsfig}
\usepackage{dcolumn}

\usepackage{graphicx}
\usepackage{bm}
\usepackage{amsmath,amssymb,amsthm}
\usepackage[colorlinks=true,linkcolor=blue,citecolor=red]{hyperref}
\newcommand{\be}{\begin{equation}}
\newcommand{\ee}{\end{equation}}
\newcommand{\bea}{\setlength\arraycolsep{2pt} \begin{eqnarray}}
\newcommand{\eea}{\end{eqnarray}}

\def\0{{\sst{(0)}}}
\def\1{{\sst{(1)}}}
\def\2{{\sst{(2)}}}
\def\3{{\sst{(3)}}}
\def\4{{\sst{(4)}}}
\def\5{{\sst{(5)}}}
\def\6{{\sst{(6)}}}
\def\7{{\sst{(7)}}}
\def\8{{\sst{(8)}}}
\def\sst#1{{\scriptscriptstyle #1}}

\makeatletter \@addtoreset{equation}{section}

\usepackage{multirow}
\begin{document}
%

\title{\normalsize
\phantom{fff}
\vspace{-3cm}
\begin{flushright}
FISPAC-TH/27/2020\\
UQBAR-TH/2020-314
\end{flushright}
\vspace{2cm}
{\bf \Large	Black Hole Shadows  in M-theory  Scenarios}}
\author{   \small A. Belhaj$^{1}$\footnote{belhajadil@fsr.ac.ma},  M.  Benali$^{1}$,  A. El Balali$^{1}$,  W. El Hadri$^{1}$,  H. El Moumni$^{2}$\thanks{hasan.elmoumni@edu.uca.ma},
E. Torrente-Lujan$^{3}$\thanks{torrente@cern.ch}
	\hspace*{-8pt} \\
	{\small $^1$ D\'{e}partement de Physique, Equipe des Sciences de la mati\`ere et du rayonnement,
		ESMaR}\\ {\small   Facult\'e des Sciences, Universit\'e Mohammed V de Rabat,  Rabat, Morocco} \\
	{\small $^{2}$  EPTHE, Physics Department, Faculty of Science,  Ibn Zohr University, Agadir, Morocco}\\
 {\small $^{3}$ IFT, Dep. de F\'isica, Univ.  de Murcia,
Campus de Espinardo, E-30100 Murcia, Spain}\\
 }

 \maketitle

	\begin{abstract}
		{\noindent}
		
		We   study  the  shadows of  four dimensional  black   holes  in M-theory inspired models.  We first   inspect the  influence of  M2-branes on such optical  aspects for  non-rotating solutions. In particular, we show  that the  M2-brane  number   can control the circular shadow size.   This      geometrical behavior     is  distorted  for  rotating solutions exhibiting cardioid shapes  in  certain moduli space regions.  Implementing a  rotation parameter, we analyze the  geometrical shadow  deformations.  Among others,  we recover the circular behaviors for a  large  M2-brane number.
  Investigating the
energy emission rate at  high energies, we   find, in a well-defined approximation,
that the associated peak  decreases with   the  M2-brane number.
 Moreover, we investigate  a possible connection with  observations (from Event Horizon Telescope or future devices)  from a  particular M-theory compactification  by  deriving  certain  constraints on the M2-brane number in the  light of
the  $M87^\star$ observational parameters. 
 \\ \\
		{\bf Keywords}:  Black holes,   Shadows,  Energy emission rate,  M2-branes, M-theory.
	\end{abstract}
\newpage	
	\tableofcontents
\section{Introduction}

Black holes were considered, until recently,  objects to be treated only by theoretical physics methods.
Nowadays,  their properties can be also observed, opening  new  windows in astrophysics, cosmology and the study of
gravity at non-trivial regimes.
Among others,  the physics of  such objects has led to remarkable results  in many gravity models including M-theory,
  a   candidate for a quantum theory of gravity \cite{Maldacena:1997de,Bah:2015nva,Gnecchi:2013mja}.

The first strong indication   of  the existence of  black holes was the observation of the
 gravitational waves \cite{Abbott:2016blz}.
  Later,  rather direct,  evidence came with the first image of the shadow of
  the black hole in $M87^\star$ galaxy announced by  the Event Horizon Telescope (EHT) international
  collaboration \cite{Akiyama:2019cqa,Akiyama:2019bqs}.
  In the near future, projects like the  Black Hole Cam (BHC)  envisage  to   take   images for Sagittarius $A^\star$, the supermassive black hole in the center of the Milky Way galaxy \cite{Goddi:2017pfy}.
  With the  $M87^\star$  black hole shadow data in hand,  we expect to make progress on the understanding of gravity models. The predictions for the  black hole physical parameters of any gravity model,  General Relativity (GR) and beyond,
    should be compatible with such
    observational data, not only at qualitative levels but  also  at increasing  levels of the  numerical accuracy.

  The EHT data  reveals that the  $M87^\star$ black hole has   a mass $M=(6.5\pm0.7)\times10^9 M_\odot$ and a
ring of  a diameter $\theta_d=42\pm3\  \mu as$ \cite{Akiyama:2019eap,Akiyama:2019fyp}.
Future improvements of the observations will allow to  have more precise and detailed information about the nature
of the black hole space-time and  strong gravity regimes.
Recently, the  exploitation of such  data has been the object  of intensive  investigations  dealing with
 the Einstein-Maxwell-Dilaton-Axion,  the  Einstein-{\AE}ther theories
 of gravity \cite{Wei:2013kza,Zhu:2019ura}, extra  dimensions \cite{Vagnozzi:2019apd}
 or GR gravity  \cite{Kumar:2020yem}.   More details and results on such activities    can be found in
\cite{Grenzebach:2014fha,Haroon:2018ryd,Johannsen:2015hib,Cunha:2015yba,Eiroa:2017uuq,Wang:2017qhh,
Tsukamoto:2017fxq,Tsupko:2017rdo,Ohgami:2016iqm,Younsi:2016azx,Abdujabbarov:2016hnw,
Amir:2016cen,Lu:2014zja,Guo:2018kis,
Bambi:2008jg,Konoplya:2019fpy,Bambi:2019tjh, li2020shadow, virbhadra2000schwarzschild, khodadi2020black,12,13}.

At low energies, M-theory   is usually described  by  an eleven dimensional supergravity
involving   solitonic brane objects. It has been shown that  this  theory can produce some non-perturbative
 limits of superstring models after certain  compactifications on particular geometries \cite{Witten:1997sc}.
 Up to   appropriate approximations,   such   models  have been linked   to  the  black hole physics  supported
 by the  AdS/CFT correspondence \cite{Maldacena:1997de}.  In this context,  a  particular emphasis has been put on
 the interplay between  such a  physics on Anti de Sitter (AdS) geometries and thermodynamics.
 This has opened ways for further research, for example the study of  links
 with black hole  thermodynamical aspects.
 Precisely, phase transitions of numerous  AdS black holes have been extensively  studied   providing
 non-trivial results \cite{Zhang:2015ova,Belhaj:2015uwa,m5brane,Belhaj:2020oun}.
 More recently,   these  thermodynamical properties  have  been  combined with  optical ones
 in order  to unveil    such  phase transitions  \cite{Belhaj:2020nqy}.
  These optical aspects, concerning  the shadow and  the deflection angle,
     have been discussed using different  approaches\cite{Wei:2013kza,belhajetal}. In particular, it has been revealed
  that the shadows of non-rotating black holes involve a circular geometry\cite{claudel2001geometry}. However, this
  geometry can be deformed by introducing the  spin rotation parameters\cite{hioki2009measurement}.   It has been  reported in multiple studies
  that  the size of such a geometry is sensitive to the physical parameters of beyond-GR models. In particular,  it is
  sensitive to the physical parameters of solutions     including  dark  sectors  \cite{konoplya2019shadow}.

The aim of this paper is the  investigation of  the shadow
 aspects  of four dimensional   black holes   in M-theory inspired models.
 We first   analyze   the   M2-brane effect  on such optical   behaviors  for  non-rotating solutions.
  In particular, we show  that the  M2-brane  number   can be used to control the circular shadow size.
   This  optical aspect   is  distorted in rotating solutions  involving  cardioid shapes for certain moduli space regions.    Precisely, we inspect the  influence of the  M2-brane number and  the rotation parameter
    on the geometrical shadow  observables.  Moreover, we discuss   a possible connection with  observations (from Event Horizon Telescope or future devices) associated with  a   particular M-theory compactification  by  imposing   certain  constraints on the M2-brane number in the  light of
the  $M87^\star$ observational parameters.
    Studying the energy emission rate at  high energies  of such M-theory black holes,
    we    find that the corresponding  peak  decreases with   the M2-brane number among other properties.

The organization of this work is as follows.
 In section 2, we investigate  certain  optical  aspects of   four dimensional non-rotating black holes in  M-theory.
 In section 3, we examine the shadow behaviors  of   the  rotating black holes  
  The energy emission rate is  analyzed in section 4.  In section 5, we attempt to give  a possible connections with with  observations.
  The last section is devoted to
  conclusions  and open questions.

\section{Shadows of  non-rotating black holes in M-theory}

We focus here on the study of the shadow properties of  a non-rotating  black hole solution
in M-theory with  M2-brane objects.
 The bosonic sector of the corresponding supergravity effective theory at low energies is given by
\begin{equation}
16 \pi G_{11} \, S = \int d^{11}x \sqrt{-h} \left( R - \frac{1}{2} \vert F_4 \vert^2 \right) - \frac{1}{6} \int A_3 \wedge F_4 \wedge F_4,
\end{equation}
where $h$ is the   space-time metric determinant and  $R$ represents  the scalar curvature.  The field strength
$F_4=dA_3$  originates from  the  gauge potential 3-form $A_3$ and $G_{11}$ is the eleven dimensional gravitational constant. A close inspection reveals that M-theory involves two brane solutions, the
 M2 and the  M5-branes.
 The near horizon geometry of such black M-branes can be written as  a  product of AdS spaces and spheres as follows
\begin{equation}
\mathcal{X}^{M-theory} = AdS_d \times \mathbb{S}^{11-d},
\end{equation}
where $d=4$ or $d=7$
corresponding to  the M2 and the  M5-branes respectively \cite{Witten:1997sc}.
In this work, we will focus on the M2, $d=4$, case  to make  contact with observational black hole  physics.  We expect  similar results   for the  M5-branes  since they are dual objects. In particular,  we deal with  a four dimensional solution relaying on  the  M2-brane physics associated with the geometry
\begin{equation}
\mathcal{X}^{M2} = AdS_4 \times \mathbb{S}^{7}.
\end{equation}
We assume  first the, non-rotating, metric
\begin{equation}
\label{ds}
ds^2=-f(r)dt^2+\frac{1}{f(r)}dr^2+r^2(d\theta^2 +\sin^2 \theta d\phi^2) +L_{AdS}^2d\Omega^2_{7},
\end{equation}
where $d\Omega^2_{7}$  indicates  the metric of  the seven dimensional real sphere with a  radius $L_{AdS}$.   In this solution, the  metric function $f(r)$  takes the form
\begin{equation}
\label{f}
f(r)=1-\frac{m}{r}+\frac{r^2}{L_{AdS}^{2}}.
\end{equation}
The integration constant $m$  is related to the  black hole mass $M$ by the following relation
\begin{equation}
\label{m}
m=2 \, G_4 \times M,
\end{equation}
where  $G_4$ is  the four dimensional gravitational constant. Since the four dimensional $AdS$  solution  is obtained from the compactification of the eleven dimensional theory on the seven dimensional sphere, one can obtain
\begin{equation}G_4=\frac{G_{11}}{\text{Vol} \left( \mathbb{S}^7 \right)},\end{equation}
where one has
 \begin{equation} G_{11}=2^4 \pi^7 \ell_p^9,  \qquad  \text{Vol} \left( \mathbb{S}^7 \right)=\omega_7 L_{AdS}^7. \end{equation}
Introducing $L_{AdS}$,  it   has been shown that the eleven dimensional space-time
\begin{equation}
\mathcal{X}^{M2}=\left( AdS_4 \right)_{L_{AdS}/2} \times \left( \mathbb{S}^{7}\right)_{L_{AdS}},\end{equation}
 described by the metric of Eq.\eqref{ds},  can be
   considered  as the near horizon geometry of $N$ coincident M2-branes in an  eleven dimensional supergravity limit of M-theory. In this way, the $AdS$ radius $L_{AdS}$ can be related  to the M2-brane  number $N$    via the  usual  relation
\begin{equation}
\label{L}
L_{AdS}^9=2^{-\frac{3}{2}} \pi^3 N^{\frac{3}{2}} \ell_p^9
\end{equation}
where $\ell_p$ is the  Planck length.
In this framework,  the mass parameter  $m$  becomes a    M2-brane  number dependent
\begin{equation}
m(N)=\frac{192\times 2^{1/6} \,\pi ^{2/3}\,\ell_p^2 \, M}{N^{7/6}}.
\label{M4}
\end{equation}
In this context, the  metric function of Eq.\eqref{f} takes  then the form
\begin{equation}
\label{ff}
f(r)= 1-\frac{192\times  2^{1/6} \,\pi ^{2/3}\,\ell_p^2\, M}{N^{7/6}\, r} + \frac{2^{1/3} \, r^2}{N^{1/3} \pi^{2/3}  \ell_p^2}.
\end{equation}

To  study  the null mass  geodesics (``photon'' orbits)  in this geometry,  in the presence of M2-branes,
we exploit   the Hamilton-Jacobi method based on the  Carter  factorization mechanism
developed in  \cite{carter1968global}. This method has been extensively used  in the rotating solutions.
Although not strictly necessary for the non-rotating case, its use will allow one  to establish a smooth contact with
the rotating case which will be  studied later on.

 According to  \cite{chandrasekhar1998mathematical},    the photon equation of motion can be  obtained    using the   Hamilton-Jacobi equation
\begin{equation}
\label{ks1}
\frac{\partial S}{\partial \tau}+\frac{1}{2}g^{\mu\nu}\frac{\partial S}{\partial x^\mu}\frac{\partial S}{\partial x^\nu}=0,
\end{equation}
 where   $\tau$ is the affine parameter along the geodesics and where  $S$  is   the Jacobi action  written as
\begin{equation}
\label{ks2}
S=-Et+L\phi+S_r(r)+S_\theta(\theta).
\end{equation}
 $E$ and $L$  are  the  conserved energy and the conserved  angular momentum, respectively. $S_r(r)$ and $S_\theta(\theta)$  are functions of $r$ and $\theta$, respectively, which  will be absorbed  in  certain relevant relations.
The geodesic equations,
 controlling  the motion of the  photon,  can be expressed as
\begin{align}
\label{Eqm1}
 \frac{dt}{d\tau}& =  \frac{E}{f(r)},\\
\label{Eqm2}
 \frac{d \phi}{d\tau}& =  \frac{L}{{r^2}\,\sin^2\theta}, \\
\label{Eqm3}
r^2 \frac{d \theta}{d\tau} &= \pm \sqrt{\Theta (\theta)}, \\
\label{Eqm4}
r^2 \frac{d r}{d\tau} &= \pm \sqrt{\mathcal{R}(r)}.
\end{align}
The expression $\mathcal{R}(r)$ and $\Theta(\theta)$    are given as a function of  the Carter constant $\mathcal{K}$ in the following way
\begin{equation}
\mathcal{R}(r)=  E^2r^4-r^2f(r)(\mathcal{K}+L^2), \quad \Theta(\theta)=\mathcal{K}-{L^2}\cot^2\theta.
\end{equation}
To analyze the shadow boundary,  the expression of the effective potential for  a radial motion  will be  exploited.  Using the  relation
\begin{equation}
\label{Veqr}
\left(\frac{dr}{d\tau}\right)^2+V_{eff}(r)=0,
\end{equation}
we  get the effective potential associated with M-theory backgrounds
\begin{equation}
\label{Veff}
V_{eff}(r)=\left[  1-\frac{192\times 2^{1/6} \,\pi ^{2/3}\,\ell_p^2\, M}{N^{7/6}\, r} + \frac{2^{1/3} \, r^2}{N^{1/3} \pi^{2/3}  \ell_p^2} \right]\frac{(\mathcal{K}+L^2)}{r^2}-E^2.
\end{equation}
The unstable circular orbits  can be obtained by maximizing the effective potential
\begin{equation}
\label{VdV}
V_{eff}\Big|_{r=r_0}=\frac{d V_{eff}}{d r}\Big|_{r=r_0}=0, \qquad  \mathcal{R}(r)\Big|_{r=r_0}=\frac{d\mathcal{R}(r)}{d r}\Big|_{r=r_0}=0,
\end{equation}
where $r_0$ represents the circular orbit radius of the photon. From   the expression of the effective potential, we arrive to the following equations
\begin{equation}
\label{VdVf}
V_{eff}|_{r=r_0}=\frac{d V_{eff}}{d r}\Big|_{r=r_0}= \left\{
    \begin{array}{ll}
        &\left[  1-\frac{192\,\pi ^{2/3} \, 2^{1/6} \,\ell_p^2\, M}{N^{7/6}\, r} + \frac{2^{1/3} \, r^2}{N^{1/3} \pi^{2/3}  \ell_p^2} \right](\mathcal{K}+L^2)-r^2\,E^2=0, \\
        \\
        & \left[ 2\, r-\frac{576\,\pi ^{2/3}\, 2^{1/6} \,\ell_p^2\, M}{N^{7/6}} \right](\mathcal{K}+L^2)=0.
    \end{array}
\right.
\end{equation}
It can be   shown  that  the  effective potential of  a  four dimensional  non-rotating black hole  involves  a   maximum for the photon sphere radius $r_0$ given by
\begin{equation}
\label{r0}
r_0=\frac{288 \times{2}^{1/6} \, \pi ^{2/3} \, \ell_p^2\, M}{N^{7/6}}.
\end{equation}
Now we are in position  to   discuss the shadow shapes in   M-theory in the absence of the  spin parameter.
For that purpose,  we define
the   impact parameters $\eta$ and $\xi$  by
\begin{equation}
\label{xe}
 \xi=\frac{L}{E}, \hspace{1.5cm}\eta=\frac{\mathcal{K}}{E^2}.
\end{equation}
With the help of  Eq.\eqref{VdVf} and  Eq.\eqref{xe}, we obtain a real   algebraic equation in terms of  $\eta$ and $\xi$
\begin{equation}
\label{sh0}
\eta+\xi^2=\frac{248832 \, \pi ^{4/3}  \,  {2^{1/3}} \, {N}^{1/3} \,\ell_p^4\, M^2}{248832 \, (2 \pi )^{2/3} \, \ell_p^2\, M^2+N^{8/3}}.
\end{equation}
To  visualize  the shadow of the  four dimensional black hole  from  M-theory,   we  shall introduce the celestial coordinates $\alpha$ and $\beta$  reported in \cite{chandrasekhar1998mathematical}.   For simplicity reasons,  we  consider  such  coordinates in the equatorial plane $\theta=\frac{\pi}{2}$. Using Eq.\eqref{Eqm1} and  Eq.\eqref{Eqm2}, one obtains
\begin{eqnarray}
\label{c1}
\alpha &= & \lim_{r_O\to \infty}(\frac{r_Op^{\phi}}{p^{t}})=-\xi,\\
\label{c2}
\beta &=& \lim_{r_O\to \infty}(\frac{r_Op^{\theta}}{p^{t}})=\pm \sqrt{\eta},
\end{eqnarray}
where $r_O$  is  the distance  between the observer and the black hole. In this way, the relation given in Eq.\eqref{sh0} becomes
\begin{equation}
\label{shf}
\alpha^2+\beta^2=\frac{248832 \, \pi ^{4/3}  \,  {2^{1/3}} \, {N}^{1/3} \,\ell_p^4\, M^2}{248832 \, (2 \pi )^{2/3} \, \ell_p^2\, M^2+N^{8/3}},
\end{equation}
showing a  perfect circular geometry.

\begin{figure}[t]
		\begin{center}
		\centering
			\begin{tabbing}
			\centering
			\hspace{8.cm}\=\kill
			\includegraphics[scale=.5]{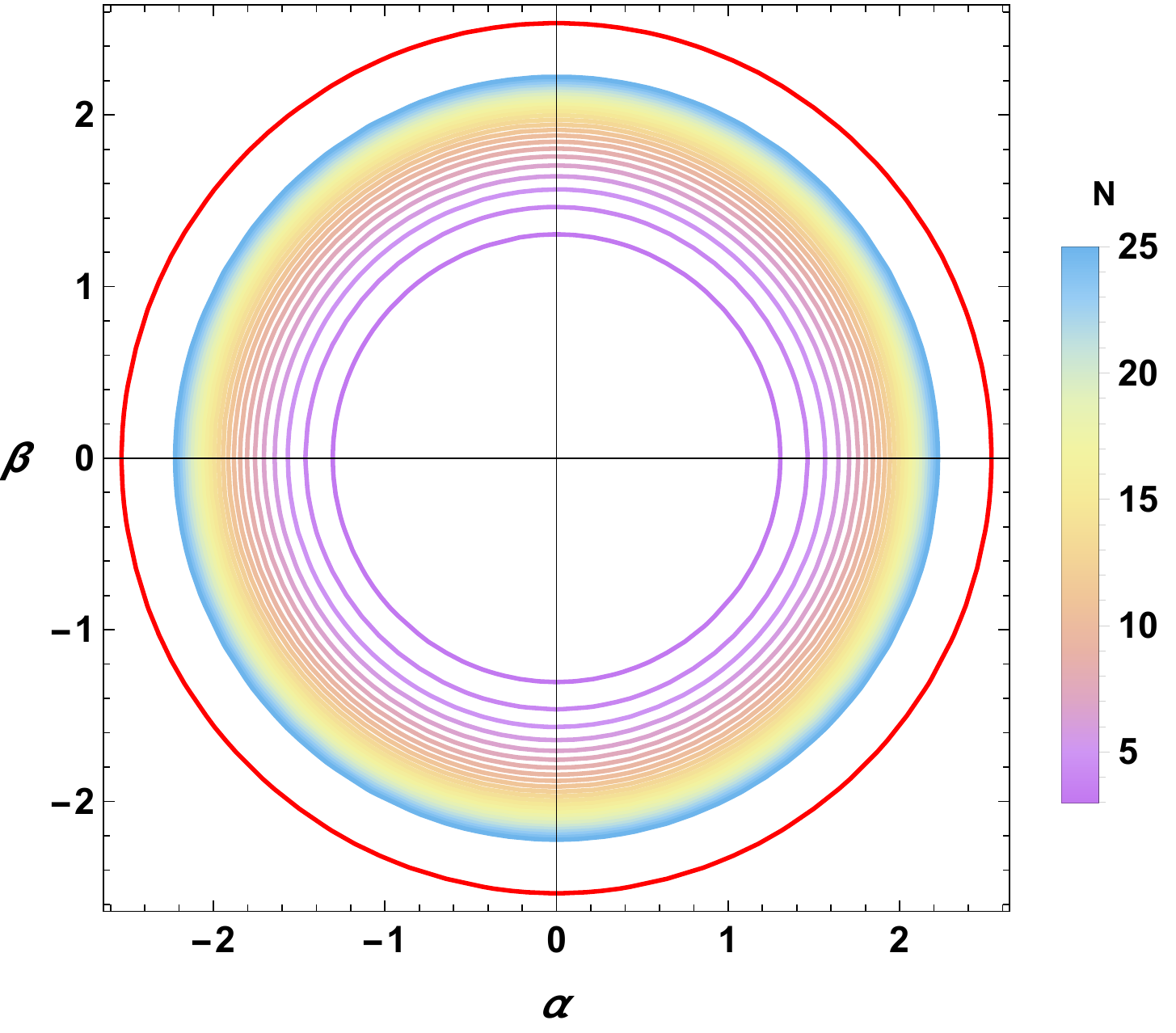} \>
			\includegraphics[width=8cm, height=6.3cm]{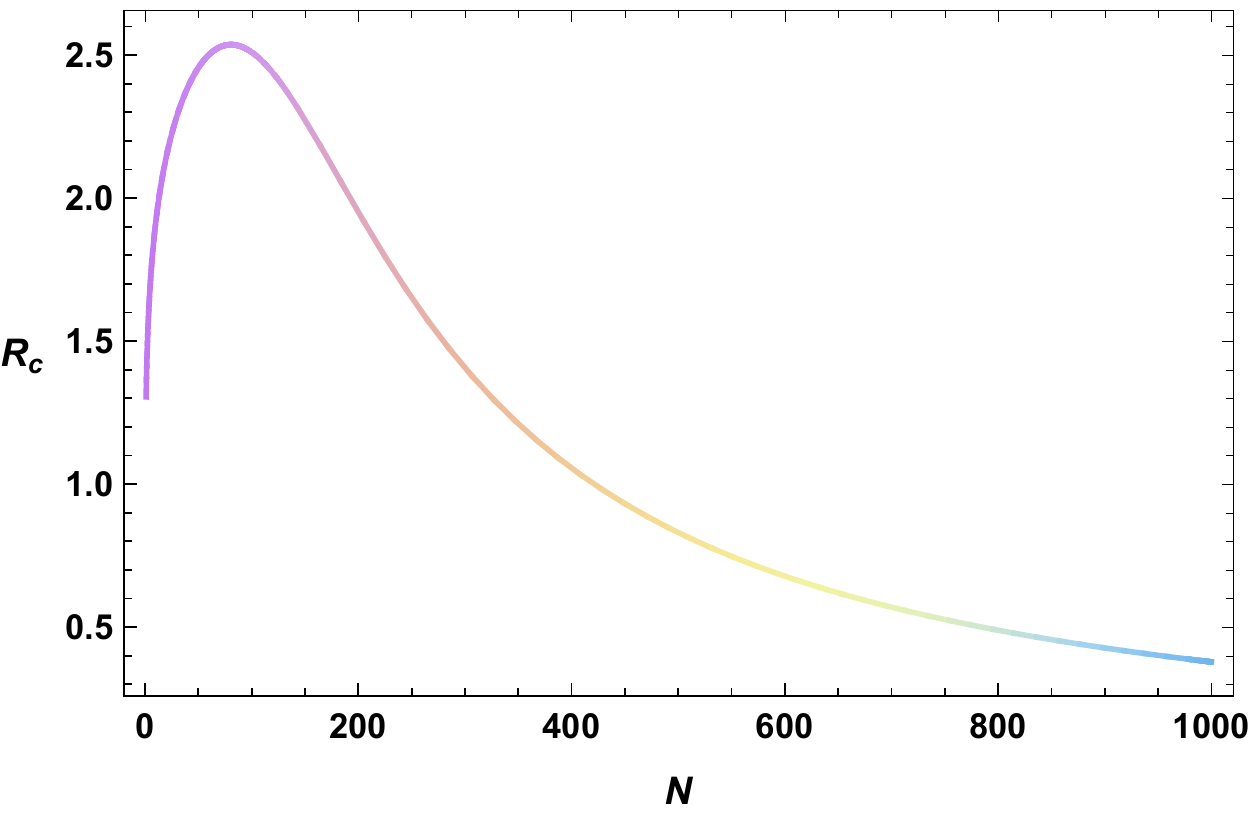} \\
		   \end{tabbing}
\caption{{
Left: Black hole shadows in the celestial plane $(\alpha-\beta)$ for  different values of the brane number $N$,  with $\ell_p=1$ and $M=1$.  The red circle corresponds to  $N= 80$.  Right: Shadow radius as a function of the brane number.}}
\label{Smsh}
\end{center}
\end{figure}

To  study  the   M2-brane  effect, we inspect   the corresponding  geometry  and the behavior  of the shadow radius  $R_c$   as a function of $N$. This is
  illustrated   in Fig.\eqref{Smsh}.

It follows  from this figure that the M2-brane number $N$, at first,  increases the circular  size of the black hole shadow
  reaching a  maximal value at
  \begin{equation}N_{max}={48 \times  7^{-3/8} \,3^{7/8} \pi^{1/4}  \,\ell_p^{3/4} M^{3/4}\,}  \end{equation}
   giving  $N_{max}\approx 80$ for $\ell_p=M=1$. Then, it  decreases  towards zero at  large values
      of $N$ $(N\rightarrow\infty)$.
 It is noted that  the gap between the circles describing the shadows decreases by increasing the number of  M2-branes.

 Having discussed the non-rotating case, we  consider next the  behaviors of  rotating M-theory black holes.  In particular, we introduce the spinning  parameter  to deal with the associated optical properties.

\section{Shadows of rotating  black holes in M-theory}

In this section, we  study    the shadow behaviors of a rotating four dimensional black hole in the presence of $N$ coincident M2-branes. Following \cite{benini2020rotating}, the line element, of the corresponding  metric solution,   reads as
\begin{equation}
ds^{2}=-\frac{\Delta_{r}}{W} \left(dt - \dfrac{a}{\Xi} \sin^2 \theta d\phi \right)^2 + W  \left(\frac{dr^2}{\Delta_{r}}+\frac{d\theta^2}{\Delta_\theta} \right) + \frac{\Delta_\theta \sin^2\theta}{W} \left( a dt -\frac{r^2+a^2}{\Xi}d\phi \right)^2,
\end{equation}
where the  involved  terms  are given by
\begin{align}
 \Delta_{r}&=r^{2}-2mr+a^{2}+\frac{r^{2}}{L_{AdS}^{2}}(r^{2}+a^{2}), ~~~~~~~~ \Delta_{\theta}=1-\frac{a^{2}}{L_{AdS}^{2}}\cos^{2}\theta, \\
 \Xi &= 1-\frac{a^{2}}{L_{AdS}^{2}},~~~~~~~~W=r^{2}+a^{2}\cos^{2}\theta.
\end{align}
To investigate  the  shadows of such  a   black hole solution,  the equations of motion will be needed. The  relevant metric    components are
\begin{align}
 g_{t \, t}&=\frac{a^2 \Delta_\theta \sin^2 \theta - \Delta_r}{W}, \\
 g_{\phi \, t} &=g_{t \, \phi}=\frac{a \sin^2 \theta \left[   \Delta_r -  \Delta_\theta (a^2+r^2) \right] }{\Xi \, W}, \\
 g_{\phi \, \phi} &=\frac{ \sin^2 \theta \left[  \left( a^2 + r^2 \right) \Delta_\theta  - a^2 \Delta_r \sin^2 \theta \right] }{\Xi^2 \, W}.
\end{align}
Using  the equation of  the canonical conjugated momentum,  we  get
\begin{equation}
\dot{t}=\frac{ E  g_{\phi \, \phi} + L g_{t \, \phi} }{\left( g_{t \, \phi} \right)^2 - g_{t \, t} g_{\phi \, \phi}}.
\end{equation}
Replacing each metric component by its expression, we obtain
\begin{equation}
W \frac{d  \, t}{d \tau} =\frac{\left(r^2 + a^2 \right)\left[ E \left(r^2 + a^2 \right)- a L \Xi  \right] }{\Delta_r} +  \frac{ a \left( L \Xi - a E \sin^2 \theta \right) }{ \Delta_\theta}.
\end{equation}
The second components of the light rays velocity  are given by
\begin{align}
 \dot{\phi}&=-\frac{E g_{t \, \phi}+L g_{t \, t}}{\left( g_{t \, \phi } \right)^2-g_{t\,t}g_{\phi \, \phi} }, \\
W \frac{d  \phi}{d \tau} &=  {\Xi}  \left[ \frac{ Ea\left(r^2 + a^2 \right) -a^2 L \Xi }{\Delta_r} + \frac{E a \sin^2 \theta - L \Xi}{\sin^2 \theta \Delta_\theta}  \right].
\end{align}
The last two components of the velocity light rays   can be  obtained by considering the Hamilton-Jacobi equation governing the geodesic motion in the  space-time.
 Using the separable variable method developed in \cite{chandrasekhar1998mathematical}, we  can get the null  geodesic equations for the photon  particle around a  rotating   black hole in the presence of $N$ coincident M2-branes. Concretely,  they  are given by
\begin{align}
  W \frac{d  \, t}{d \tau}& =  E \left[ \frac{\left(r^2 + a^2 \right)\left[ \left(r^2 + a^2 \right)- a \xi \Xi  \right] }{\Delta_r} +  \frac{ a \left(\xi \Xi - a  \sin^2 \theta \right) }{ \Delta_\theta} \right], \\
W  \frac{d  \, r}{d \tau} &=\pm\sqrt{\mathcal{R}(r}),\\
 W\frac{d  \, \theta}{d \tau} & =\pm\sqrt{\Theta(\theta)},\\
W \frac{d  \, \phi}{d \tau} &= E \;\Xi \left[ \frac{a\left(r^2 + a^2 \right) -a^2 \xi \Xi}{\Delta_r} + \frac{a \sin^2 \theta - \xi \Xi}{\sin^2 \theta \Delta_\theta}  \right].
\end{align}
In the  M-theory rotating case, the  expressions of $\mathcal{R}(r)$ and ${\Theta}(\theta)$  are found to be
 \begin{align}
\mathcal{R}(r)&=E^2\left[ \left[ \left( r^2 + a^2 \right)  -a \xi \Xi  \right]^2- \Delta_r \big[\frac{\left(a - \xi \Xi  \right)^2}{\Delta_{\theta}}+ \eta \big]   \right],\\
 \Theta(\theta)&= E^2 \left[ \eta \Delta_\theta  - \cos^2 \theta \left( \frac{\xi^2 \Xi^2}{\sin^2 \theta} - a^2  \right) \right].
\end{align}
As in the non-rotating solution, the  unstable circular orbits correspond to the maximum value of the effective potential.   Roughly,  a generic solution of such  impact parameters can be given in terms of $r_0$, $N$, $a$ and $\theta$ being  needed for making contacts with other results including the observational ones. In fact, they are given by
\begin{align}
& \xi=\frac{ \left(r^2+ a^2\right) \Delta_r^\prime-4 r \Delta_r }{a\;
\Xi\; \Delta_r^\prime} \bigg\vert_{r=r_0},\\
& \eta = \frac{ r^2  \left( 16\;  a^2\; \Delta_r\;\Delta_\theta-\left(r \Delta_r^\prime
-4 \Delta_r\right)^2 \right)}{a^2 \;  { \Delta_r^\prime}^2 \;\Delta_\theta}
\bigg\vert_{r=r_0} ,
\end{align}
where one has used  $\Delta_r^\prime=\frac{\partial \Delta_r}{\partial r}$. Precisely, an appropriate calculation generates the following relations
\begin{align}
& \xi=\frac{a^2 \left(\mu -\nu  r_0^3+r_0\right)+r_0^2 (r_0-3 \mu )-a^4 \nu r_0}{a \left(a^2 \nu -1\right) \left(a^2 \nu  r_0-\mu +2 \nu  r_0^3+r_0\right)},\\
& \eta = \delta  +\frac{4 a^4 \nu  r_0^2 \cos ^2(\theta ) \left(a^2 \left(\nu  r_0^2+1\right)+r_0 \left(-2 \mu +\nu  r_0^3+r_0\right)\right)}{a^2 \left(a^2 \nu  \cos ^2(\theta )-1\right) \left(a^2 \nu  r_0-\mu +2 \nu  r_0^3+r_0\right)^2}
\end{align}
 where   $\delta$, $\mu$, and $\nu$ are expressed as follows
\begin{align}
\delta &=\frac{r_0^3 \left(4 a^2 \mu +6 r_0^2 \left(a^2 \mu  \nu +\mu \right)-r_0^3 \left(a^2 \nu -1\right)^2-9 \mu ^2 r_0\right)}{a^2 \left(a^2 \nu  \cos ^2(\theta )-1\right) \left(a^2 \nu  r_0-\mu +2 \nu  r_0^3+r_0\right)^2},\\
\mu &=\frac{192 \times 2^{1/6} \,\pi ^{2/3}\,\ell_p^2\, M}{N^{7/6}}, \\
\nu& = \frac{2^{1/3}}{N^{1/3} \pi^{2/3}  \ell_p^2}.
\end{align}
Now, we   analyze the shadow behaviors of a particular situation.  Placing an  observer at infinity and considering  the photons (null mass)   near the equatorial plane $\theta=\frac{\pi}{2}$, we obtain the impact parameters
\begin{eqnarray}
\label{Xi}
\xi&=&\frac{a^2 \left(\mu -\nu  r_0^3+r_0\right)+r_0^2 (r_0-3 \mu )-a^4 \nu r_0}{a \left(a^2 \nu -1\right) \left(a^2 \nu  r_0-\mu +2 \nu  r_0^3+r_0\right)},\\
\label{Eta}
\eta&=&\frac{r_0^3 \left(4 a^2 \mu +6 r_0^2 \left(a^2 \mu  \nu +\mu \right)-r_0^3 \left(a^2 \nu -1\right)^2-9 \mu ^2 r_0\right)}{a^2 \left(a^2 \nu  r_0-\mu +2 \nu  r_0^3+r_0\right)^2}.
\end{eqnarray}
Following  \cite{chandrasekhar1998mathematical}, the   celestial coordinates in the equatorial plane given by Eq.\eqref{c1} and  Eq.\eqref{c2}  are modified  in the rotating case as follows
\begin{eqnarray}
{\alpha }& = & -\frac{\xi}{ \sin \theta_0}, \\
{\beta } & = & \pm \sqrt{\eta +a^2\cos^2\theta_0 -\xi^2 \cot^2 \theta_0},
\end{eqnarray}
where $\theta_0$ is the  inclination angle.

Using the expression of Eqs.\eqref{Xi}-\eqref{Eta}, and the celestial coordinates in the equatorial plane, we can
 illustrate the shadows of such a   four dimensional rotating    black hole for certain regions of the moduli space.

\begin{figure}[ht]
		\begin{center}
		\centering
			\begin{tabbing}
			\centering
			\hspace{8.6cm}\=\kill
			\includegraphics[scale=.5]{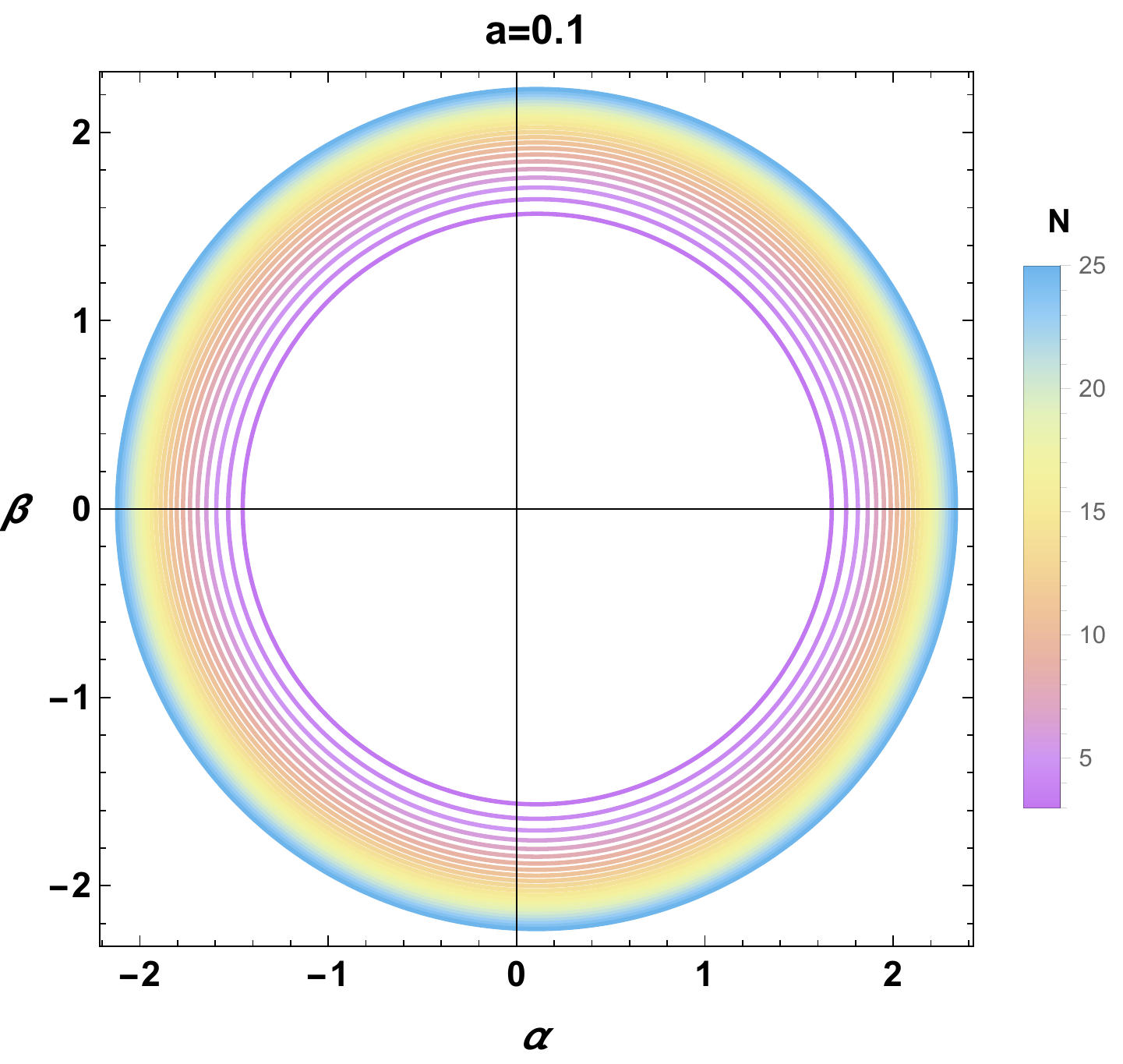} \>
			\includegraphics[scale=.5]{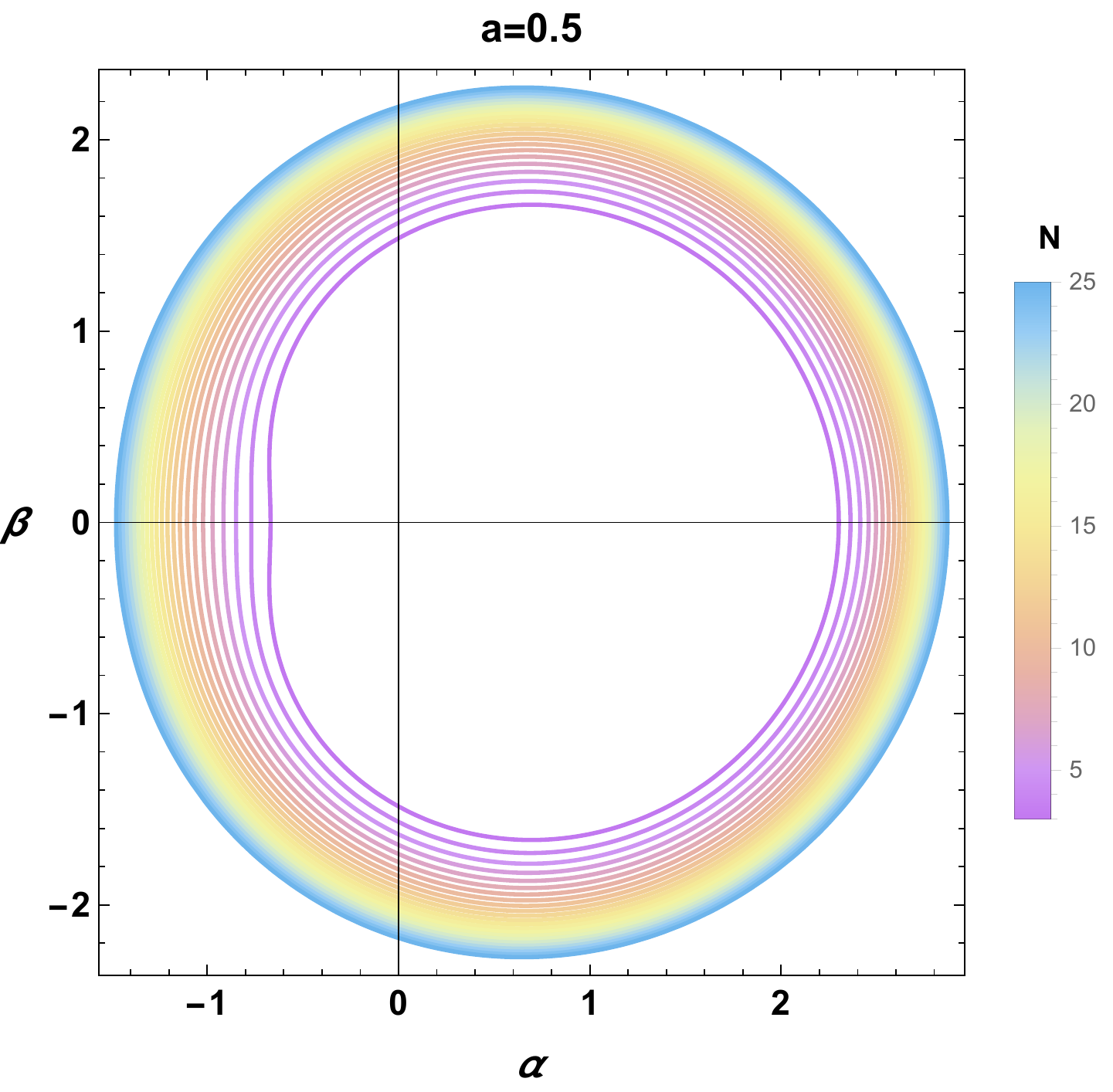} \\
			\includegraphics[scale=.5]{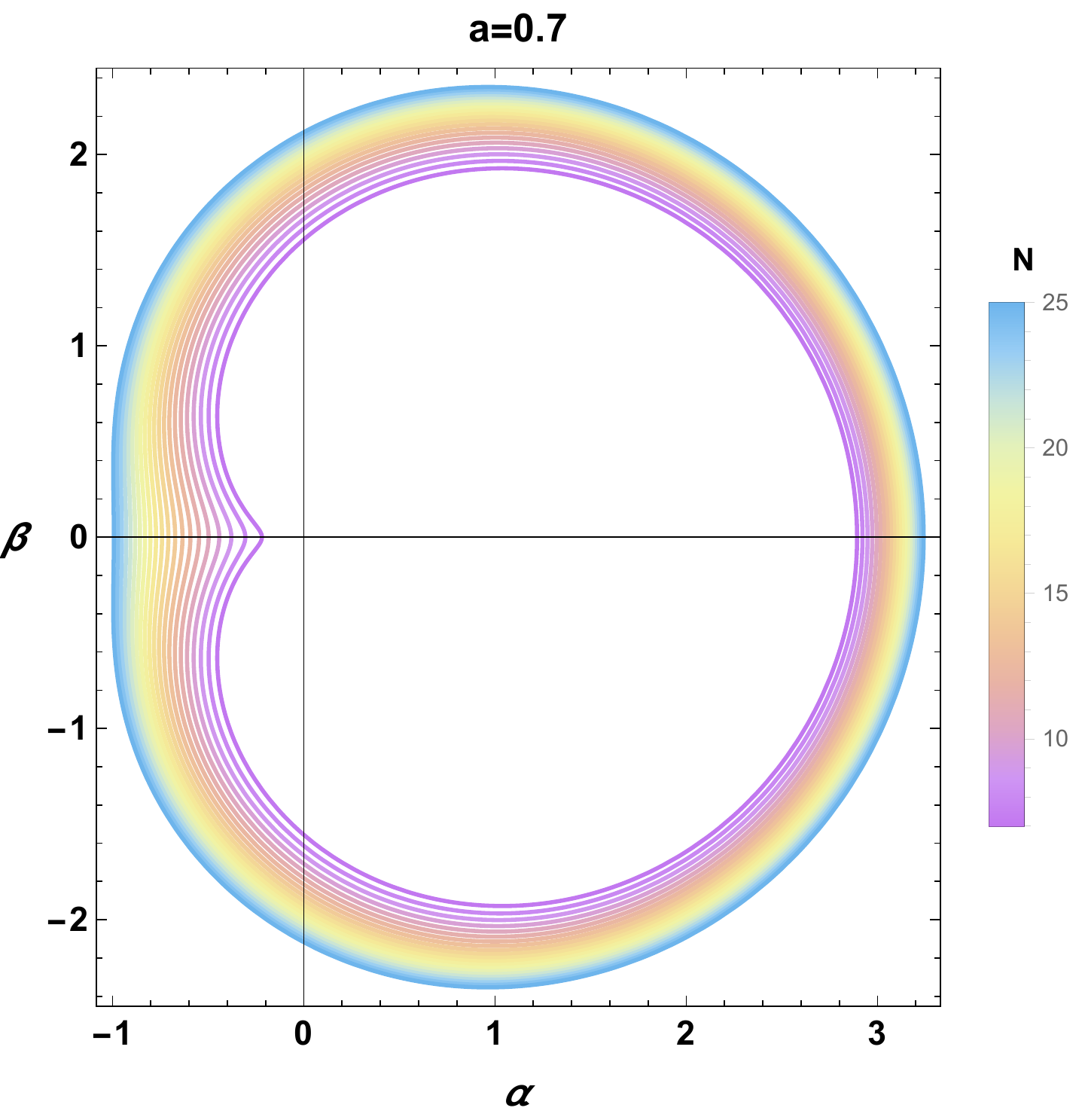} \>
			\includegraphics[scale=.5]{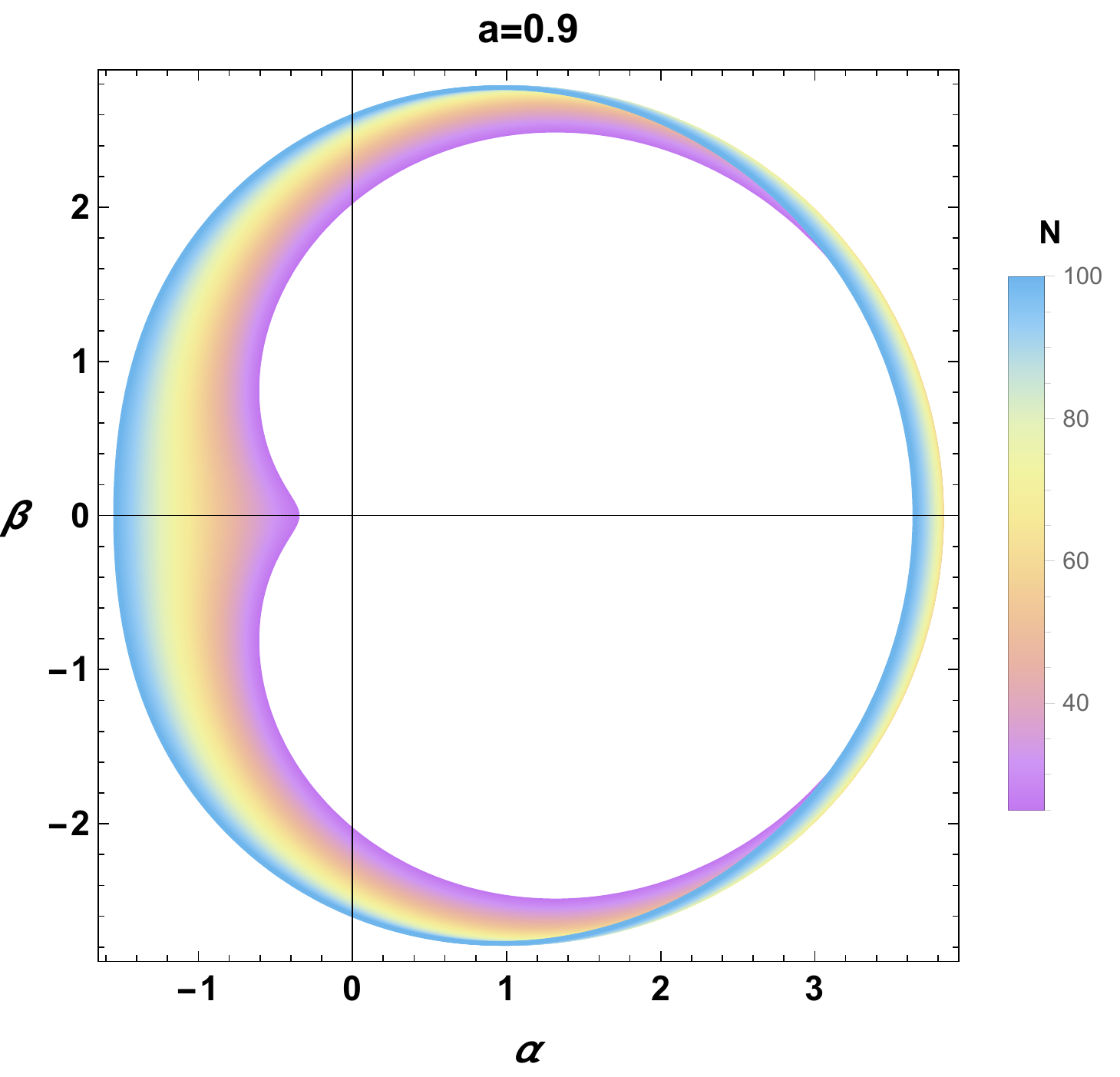}
		   \end{tabbing}
\caption{{
Shadow behaviors of a rotating black hole for different values of $M2$-brane number $N$ and the rotating rate $a$, for $\ell_p=1$ and $M=1$.}}
\label{shfa}
\end{center}
\end{figure}

 In what follows, we   discuss   the effect of
 the parameter $a$  on M-theory shadow behaviors. It is worth noting    that we recover the circular geometry of the non-rotating case from  very small
 values of  $a$, turning off   the geometrical spinning effect.
 Using  the  real solutions of    Eqs.\eqref{Xi}-\eqref{Eta}, we
 plot  in Fig.\eqref{shfa}    the associated shapes for different values of the M2-brane number $N$ and the rotation rate $a$.

We observe from  this figure that for small   values of the rotation rate $a$,  the shadows    maintain the same  circular geometries   being slightly unsymmetrical about  the central axes. In such a region of the  moduli  space,  the  M2-brane number plays the same role as in  the non-rotating case by controlling the shadow size. Increasing the value of the rotation rate $a$,  the shadows  are  shifted from the center providing a  D-shape. However,  such a   D-shape changes to a  cardioid geometry   by decreasing the number $N$ of  the M2-branes.   This  behavior has been obtained from other  effects  including  the   stringy and  the dark matter sectors  \cite{jusufi2019black,our}. It would be interesting to  come back to this remark in  future works. It is noted    that when we increase the brane number, such a shape becomes circular again showing that the brane number effect can oppose to the rotation one.

\begin{figure}[tb]
		\begin{center}
		\centering
			\begin{tabbing}
			\centering
			\hspace{8.4cm}\=\kill
			\includegraphics[scale=.5]{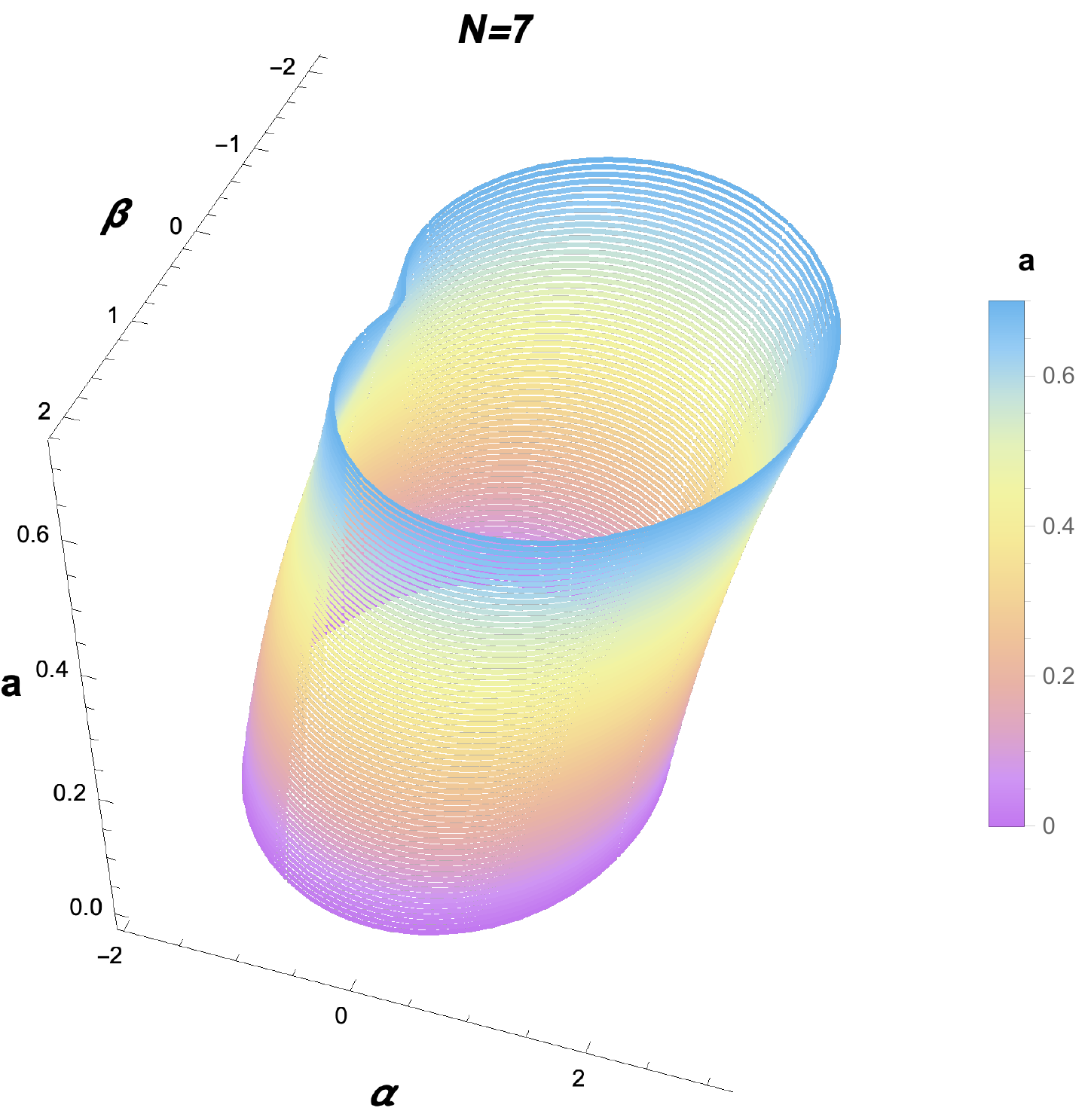} \>
			\includegraphics[scale=0.5]{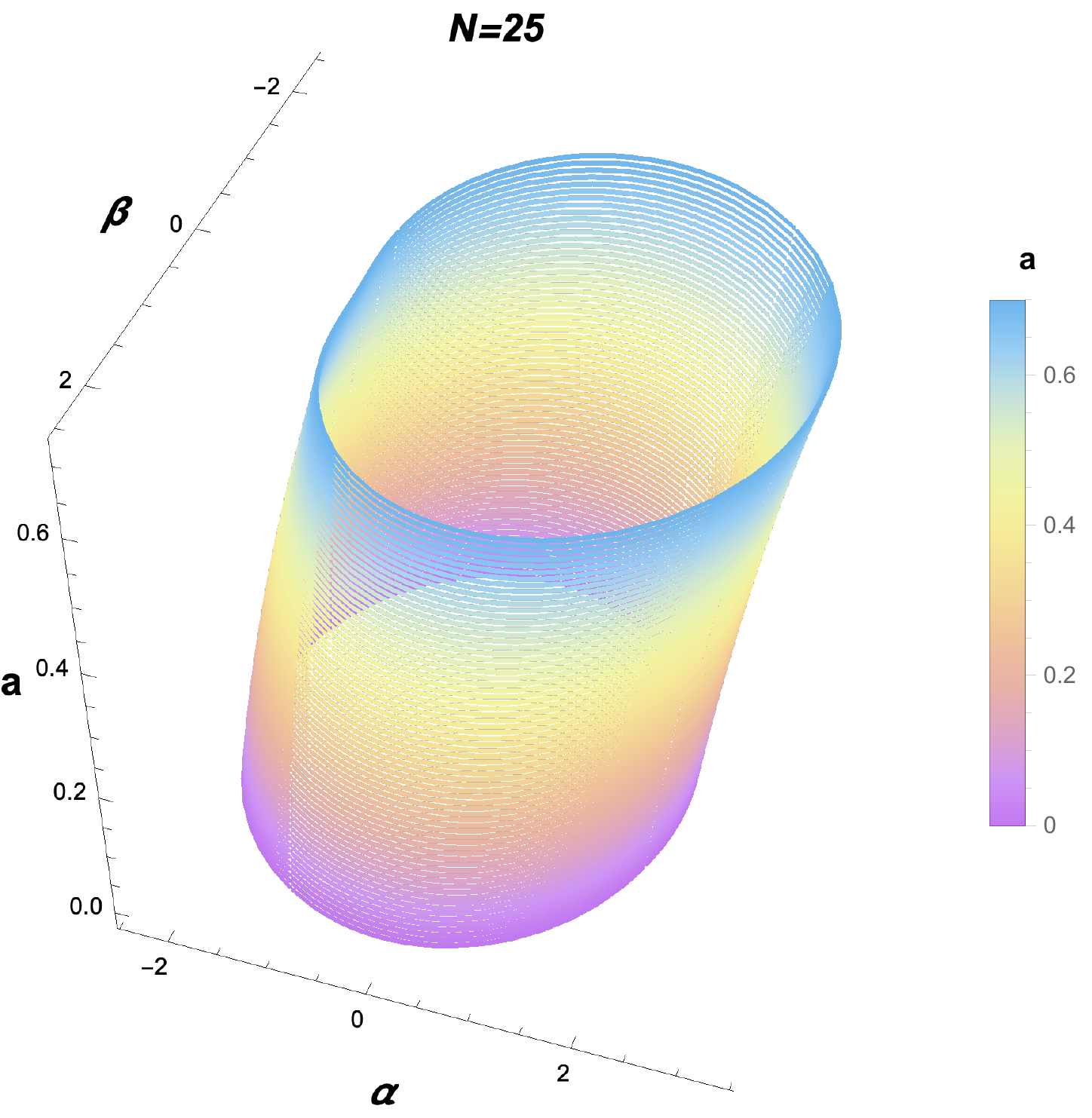}
		   \end{tabbing}
\caption{{
  Shadow behaviors  inspired by  geometric fibrations for  $N=7$ and $N=25$  with  $\ell_p=1$ and $M=1$ .}}
\label{enemroo}
\end{center}
\end{figure}

It is useful to   change the role of  $a$  from
a parameter to  a variable. This will  reduce the associated  moduli space  by considering   the
M2-brane number as  a relevant parameter.  In this regarding,   the shadow geometry will be
embedded  in a three dimensional space, instead of the two dimensional one. For $N\leq 25$, these  geometric
configurations   are  illustrated in Fig.\eqref{enemroo}. As expected,  the  M2-brane number is  confirmed
as  a real size parameter  of the  shadow fiber   over the interval ($[0,0.7]$)  corresponding to  the  spinning
coordinate. In particular,  we  remark  that the size of the shadow fiber augments over  the  segment base.  Inspecting  the   M2-brane effect,  we realize  that  two behaviors could appear. For small  values of such a parameter, the shadow fiber develops D and cardioid shapes. However, we recover circular  fibers  by augmenting   the  M2-brane number. For  a very  large number of M2-branes, the fiber size  starts to decreases until it  shrinks to zero.

\label{f2}

\begin{figure}[hb]
		\begin{center}
		\centering
			\begin{tabbing}
			\centering
			\hspace{8.4cm}\=\kill
			\includegraphics[scale=.5]{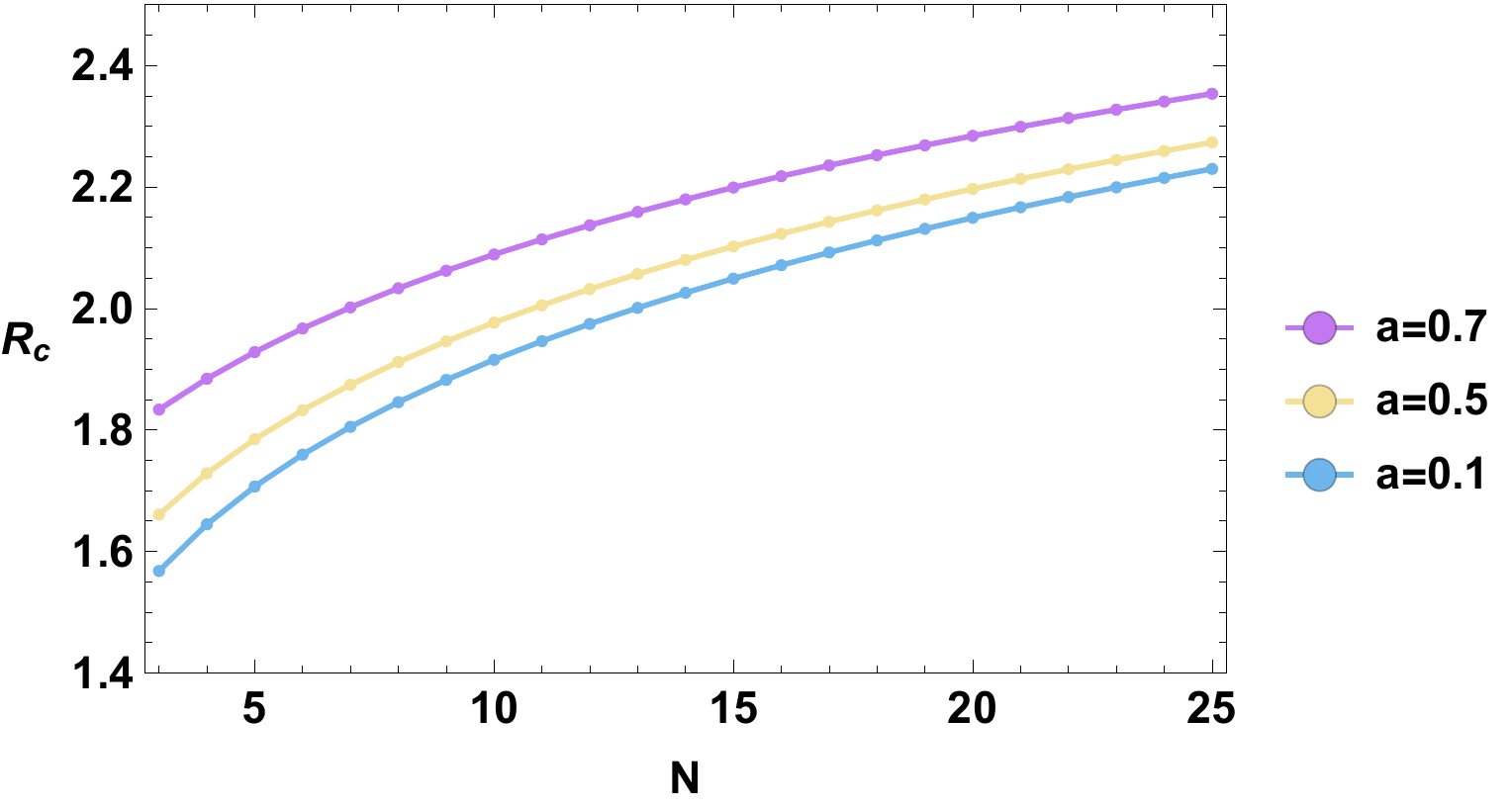} \>
			\includegraphics[scale=0.5]{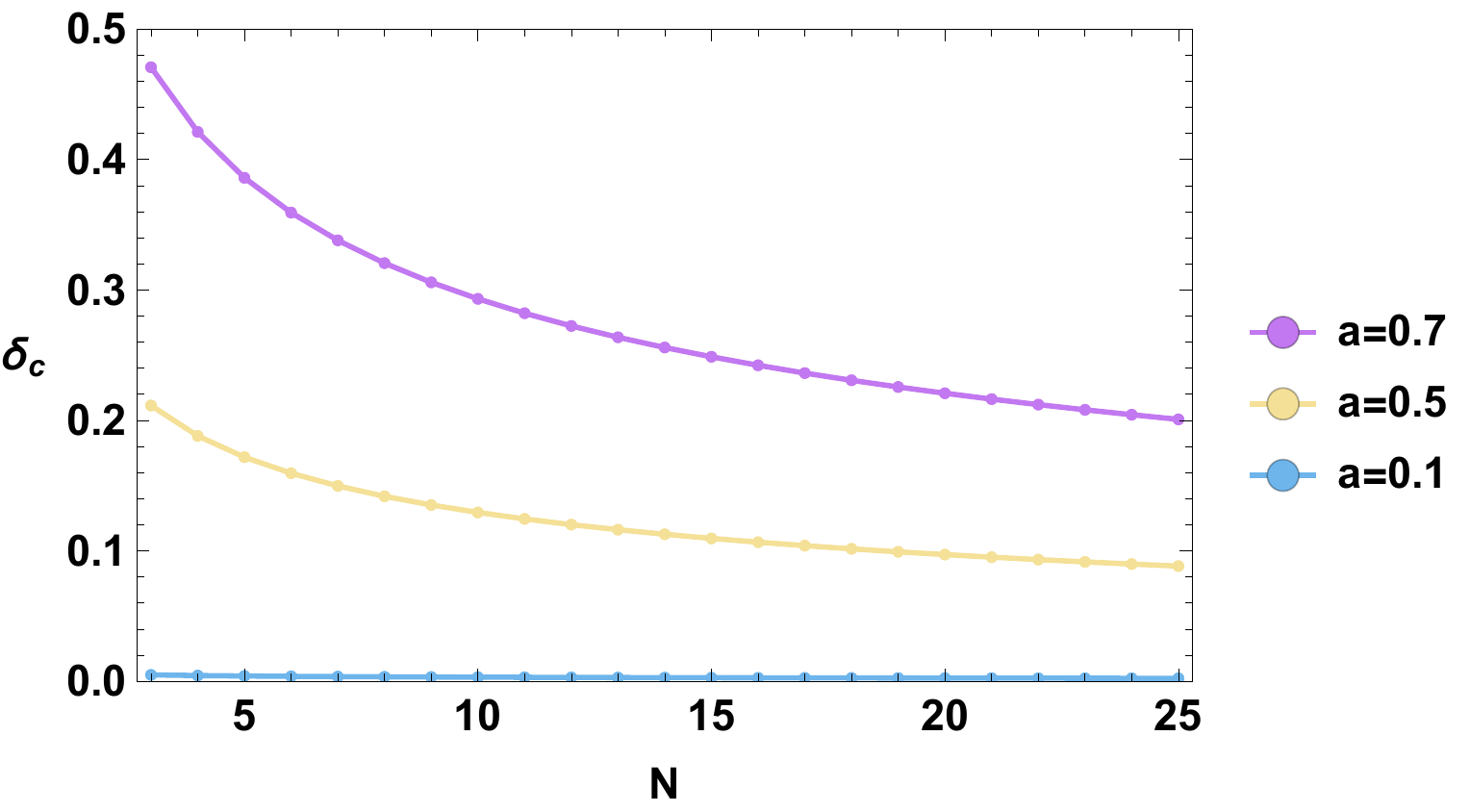}
		   \end{tabbing}
\caption{{
Astronomical observables $R_c$ and $\delta_c$, for different values of  $N$ and  $a$  with  $\ell_p=1$ and $M=1$.}}
\label{diso}
\end{center}
\end{figure}

We approach now the study of observables
which can be useful for  making contact with other results concerning  the  rotating black holes.
Two astronomical observables $R_c$ and $\delta_c$ controlling  the size and  the shape of the shadows, respectively,
will be considered.  Following   \cite{hioki2009measurement,amir2016shapes},
 the size  is  defined  by three specific points which
are the  top and  the bottom position of shadows ($\alpha_t,\beta_t$), ($\alpha_b,\beta_b$) and
a reference point ($\tilde{\alpha}_p,0$). The point of the distorted shadow circle ($\alpha_p,0$)    meets the horizontal axis  at $\alpha_p$.   In this way, the  distance between the two  latter points   is    given in terms of  a  diameter parameter
\begin{equation}D_c= \tilde{\alpha}_p- {\alpha}_p=2R_c-(\alpha_r-\alpha_p).\end{equation}
In this  geometrical  picture,   the  parameter $R_c$ of  the shadow   is approximatively  given by
\begin{equation}
\label{d1}
R_c\simeq\frac{(\alpha_t-\alpha_r)^2+\beta_t^2}{2|\alpha_t-\alpha_r|}.
\end{equation}
The  distortion parameter  is characterized by  the ratio  of  $D_c$ and $R_c$
\begin{equation}
\label{ }
\delta_c=\frac{D_c}{R_c}.
\end{equation}
It is straightforward to find analytic expressions for the theoretical predictions of such quantities in terms of
microscopical parameters.   We plot  the results in
Fig.\eqref{diso}  where we  present $R_c$ and $\delta$ as a functions of the rotation parameter $a$ and the  M2-brane number $N$.
It is observed  that, for a fixed rotation rate $a$,  the shadow size increases for large brane number values $N$. The same behavior appears  by fixing  $N$  and increasing the  spinning parameter $a$.
Besides,   the shadow size  is larger for $a\simeq 0.99$ in comparison with other values of the parameter $a$.
Moreover,  the distortion parameter
 $\delta_c$,  at fixed  values of the rotating parameter $a$, decreases  for large  values of  $N$.
 For a fixed brane number $N$,  however, the distortion parameter $\delta_c$  increases when the parameter $a$ does. Taking $a=0.1$, it has been  shown that the distortion is almost null   providing a circular geometry. A similar behavior is recovered in terms of $N$.
 For values bigger than  $N=100$,  for instance,   the distortion  becomes  very small implying that  one can recover the
 perfect  circular geometry.

\section{Energy emission rate}
It  is recalled that, for a far distant observer and very high energies, the absorption cross-section approaches the
geometrical optical limit associated with  the shadow of the black hole. In other regimes,  the absorption cross-section oscillates around
this geometrical limit.  This  limit  has been expressed in terms of the geodesic properties for   various  theories.    It has been considered as the geometrical cross section of the photon
sphere. Based on this approximation, it  takes the following form
\begin{equation}
\label{emir}
\frac{d^2 E({\omega})}{d{\omega} dt}=\frac{2\pi^{3} R_c^{2}}{e^{\frac{{\omega}}{T_i}}-1} {\omega}^3,
\end{equation}
where ${\omega}$ is the emission frequency and  $R_c$ is the shadow radius  \cite{wei2013observing}.  Here, $T_i$ denotes  the temperature of the involved   black hole   given  in terms of  the horizon radius $r_h$.

\subsection{Non-rotating   case}
\begin{figure}[!ht]
		\begin{center}
			\includegraphics[scale=0.6]{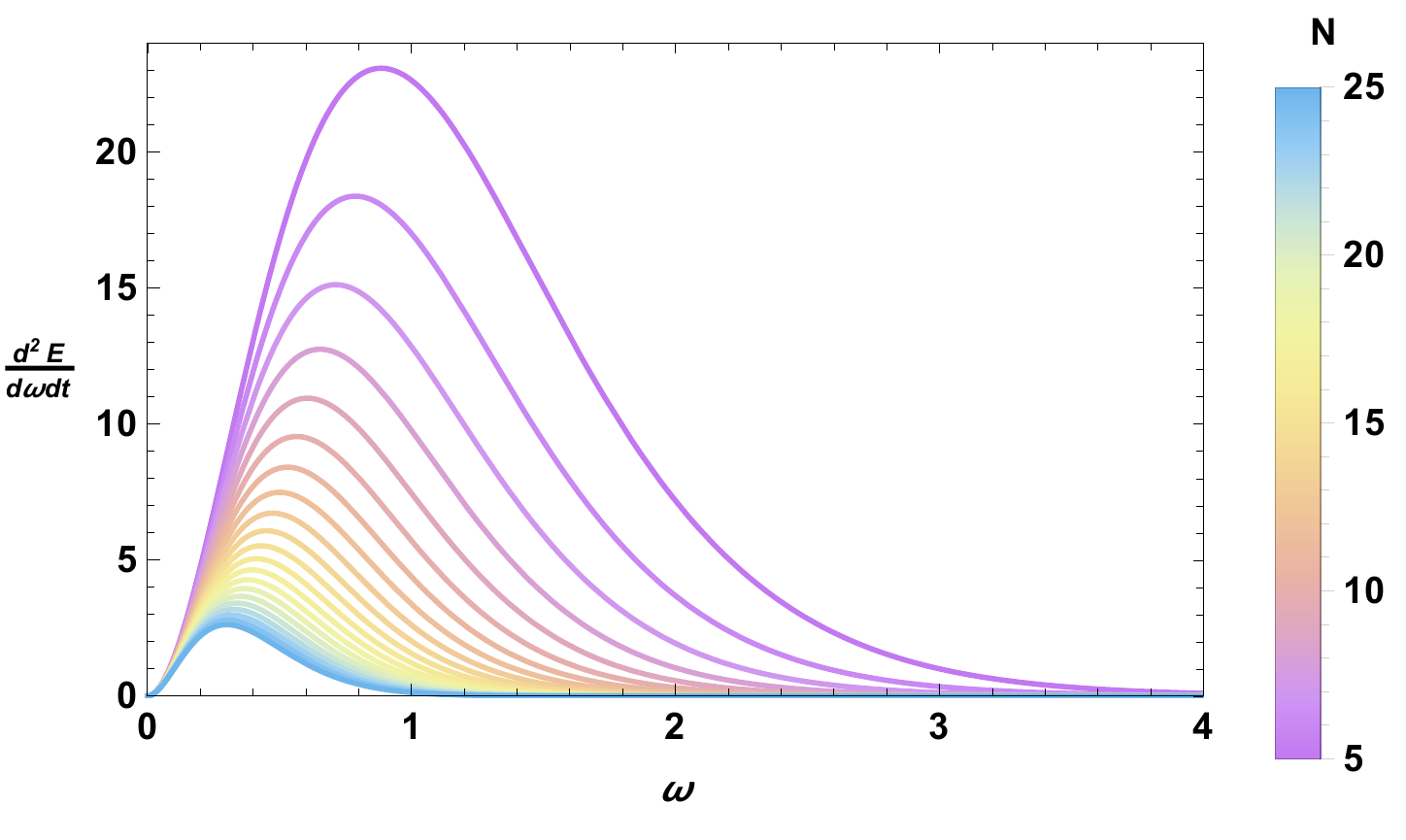}
\caption{{
Energy emission rate of non-rotating black holes as a function of the emission frequency for different brane number values, with  $\ell_p=1$ and $M=1$.}
}\label{emira}
		\end{center}
  \end{figure}

We first  consider  the non-rotating solution dealt with in  the second section.   For this solution, the temperature is expressed as
\begin{equation}
T_{nr}= \frac{1}{2 \pi  r_h} \left( 1  -\frac{192 \, \pi ^{2/3} \,2^{1/6} \ell^2_p \,M}{N^{7/6} \,r_h}+\frac{2^{4/3} \,r_h^2}{\pi ^{2/3}\, \ell_p^2 \,N^{7/6}}\right).
\end{equation}
The energy emission rate is plotted  in Fig.\eqref{emira} as a function of ${\omega}$ for  different numbers of M2-branes.

It is  remarked  from Fig.\eqref{emira} that, despite the M2-brane number $N$, the energy emission rate varies  slowly.  This   indicates  that  the evaporation process for this M-theory  black hole  can be very slow. Increasing the M2-brane number,  we get  a lower energy emission rate. This  shows that the M2-brane number could be considered as  a useful
tool  to tune  the black hole stabilization. For $N \geq 18$,  it  has been  observed    that  the energy emission rate  decreases very slowly. For  $N\geqslant N_{max} $, however,  the   energy emission rate is almost zero.

\subsection{Rotating case}

\begin{figure}[!htb]
		\begin{center}
		\centering
			\begin{tabbing}
			\centering
			\hspace{8.7cm}\=\kill
			\includegraphics[scale=.5]{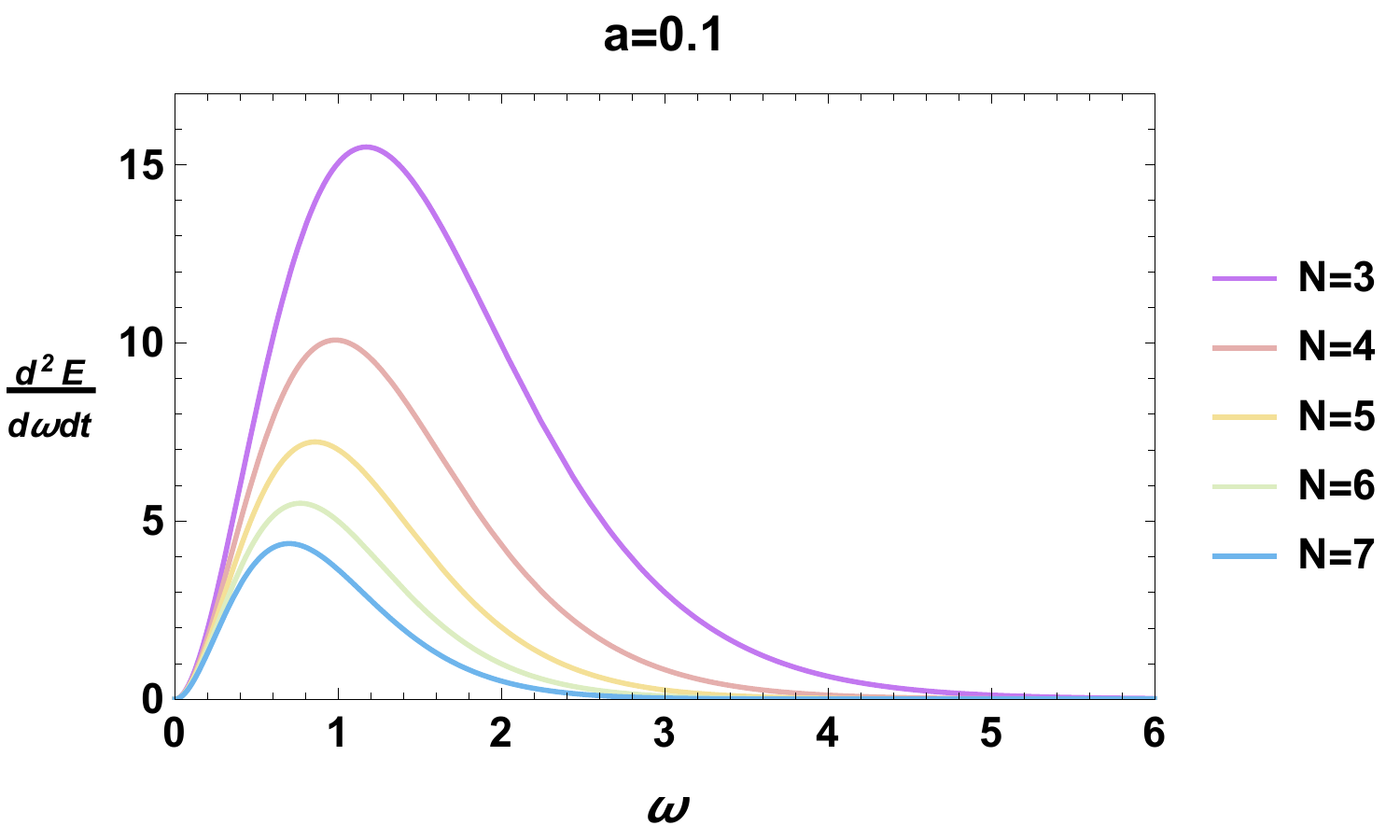} \>
			\includegraphics[scale=0.5]{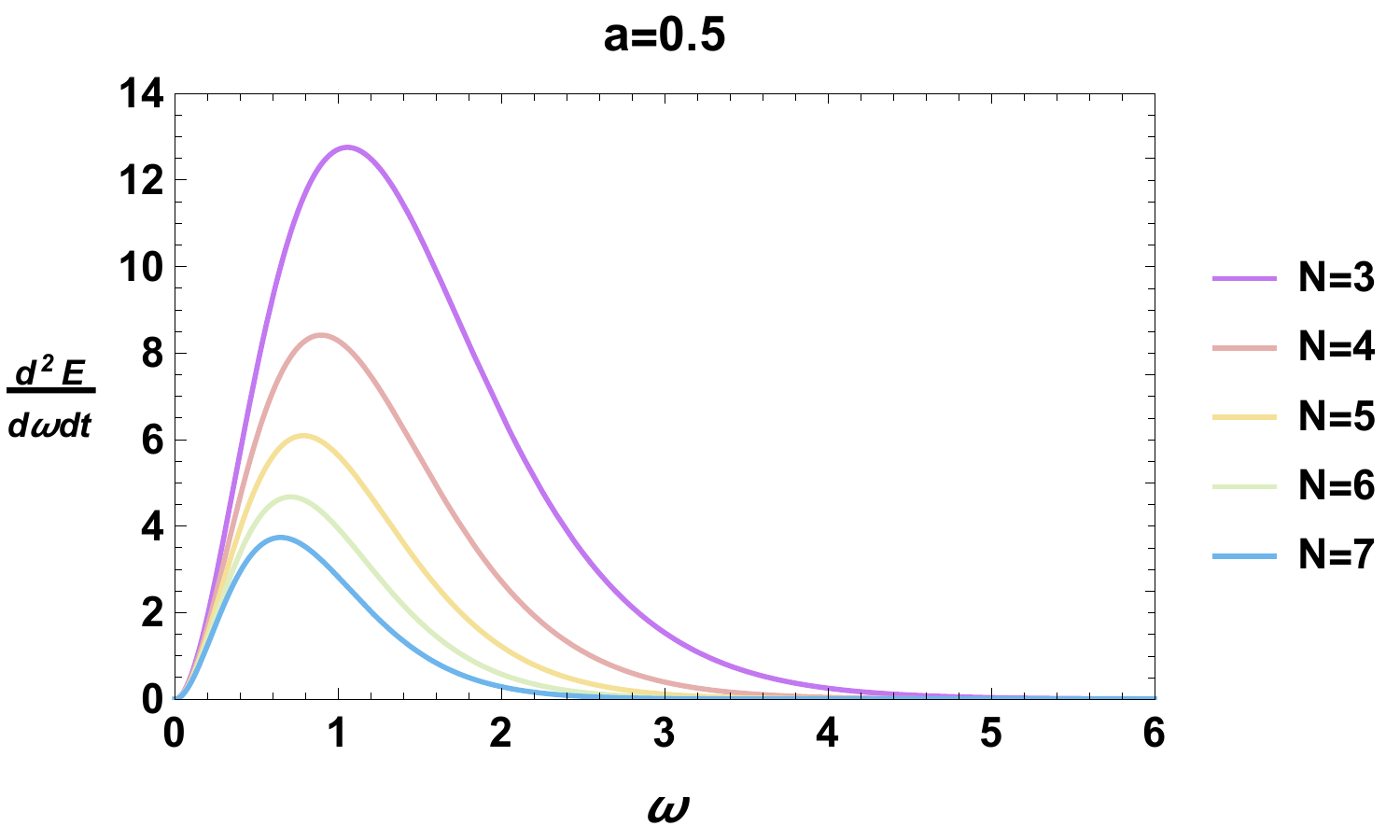} \\
			\includegraphics[scale=0.5]{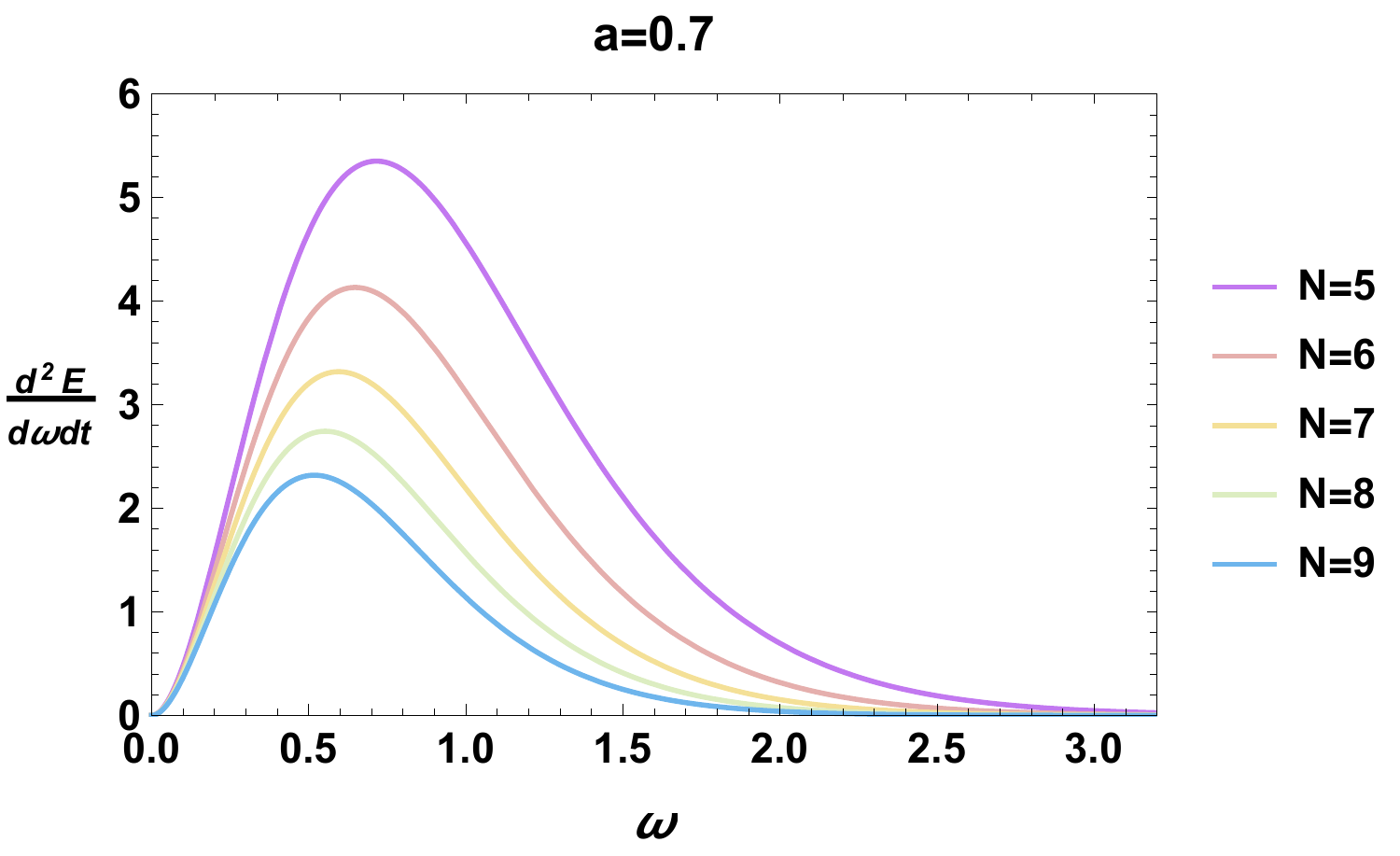} \>
		    \includegraphics[scale=0.5]{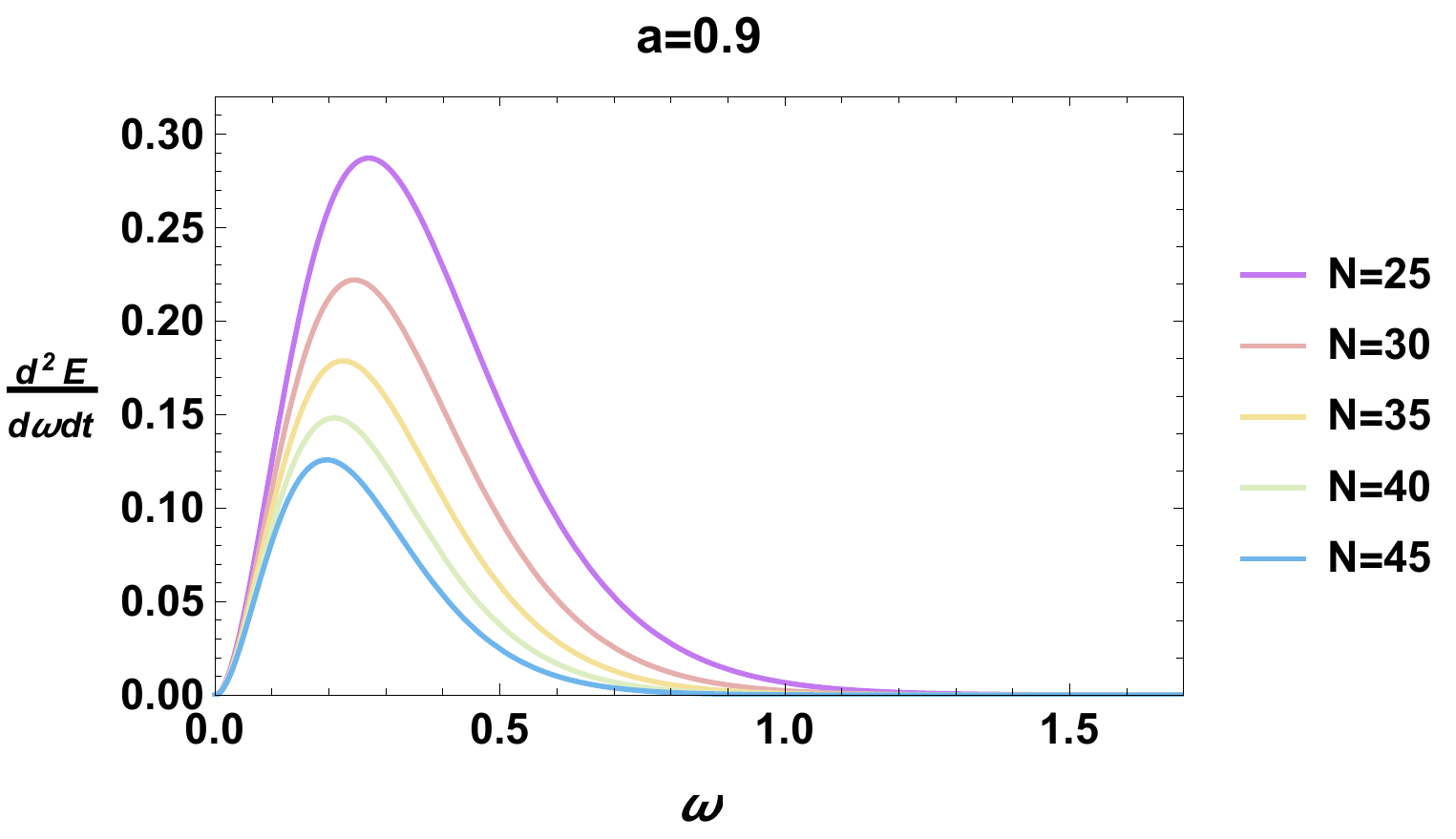}
		   \end{tabbing}
		   \caption{{{
		   Energy emission rate of rotating black holes  as a function of the emission frequency for different values of  $N$ and  $a$, with  $\ell_p=1$  and $M=1$.}}}
\label{enemro}
\end{center}
\end{figure}

 In the rotating case, using similar  computations,  we  should  get  the  temperature expressed   in terms of  the horizon radius $r_h$. Indeed, it is  found to be given by
\begin{equation}
T_{r}= \frac{r_h }{2 \pi  \,\left(a^2\,+\,r_h^2\, \right)} \left(1\,+\, \frac{2^{1/3} \,\left(a^2\,+\,2 r_h^2\,\right)}{\pi ^{2/3}\, \ell_p^2\, N^{1/3}}-\frac{192 \,\pi ^{2/3} \,2^{1/6} \,\ell_p^2\, M}{N^{7/6} \,r_h}\right).
\end{equation}
As in the previous model case,  we illustrate in Fig.\eqref{enemro} the energy emission rate  as a function of the emission frequency ${\omega}$ for  different numbers of the M2-branes and the  rotation rate $a$.

We observe, from  this  figure,  a  non-trivial behavior in the rotating case.
The particle emission rate increases by decreasing the M2-brane number.
The evaporation process is clearly faster compared to the non-rotating solution.
Increasing the rotation
rate parameter, we remark that the emission frequency as well as the energy emission rate are
lowered.
In this way, we obtain  similar  effects  on the energy emission rate in terms of the
M2-brane number as the non-rotating case.
Moreover,   we find that the
energy emission values are quite high compared to the other cases.  Increasing the parameter $a$,
 one needs more branes to get  reasonable     results.
This can be converted into a constraint on the M2-brane number $N$ provided by  the emission rate
reality condition,   matching with   the rotating black hole  properties.

\section{ Observational constraints in the light of  the $M87^\star$ image}

 The observation of the shadow of the supermassive black hole  $M87^\star$,  obtained by  EHT  collaboration,  has provided many promoting roads to probe
ceratin  gravity regimes and alternative physical  theories.  For such  reasons,  it  should be interesting to make contact with such activities. Indeed, the observational data  can put some  constraints on the relevant  black hole parameters.
In this  way, the shadow geometrical behaviors  can be visualized in terms of the brane number for different values of rotation parameter $a$.   In the unit of $M87^\star$ mass,  we superpose  the shadow of the $M87^\star$ black hole  modeled by the  a Kerr solution (red line) and the shadow of  the present M-theory model (cyan line).  For different  values of $a$,  this behavior is  illustrated   in  Fig\ref{f2}.
\begin{figure}[ht!]
		\centering
			\begin{tabbing}
			\hspace{5.2cm}\= \hspace{5.2cm}\=\kill
			\includegraphics[scale=.5]{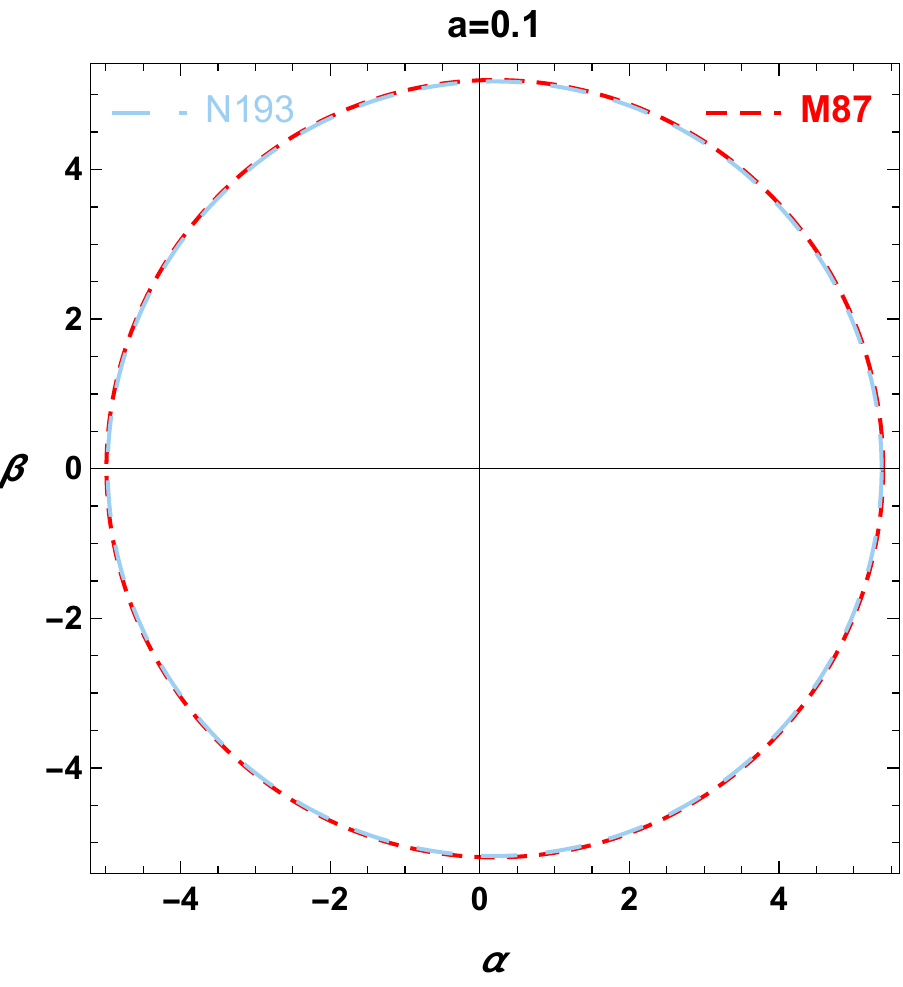} \>
			\includegraphics[scale=.49]{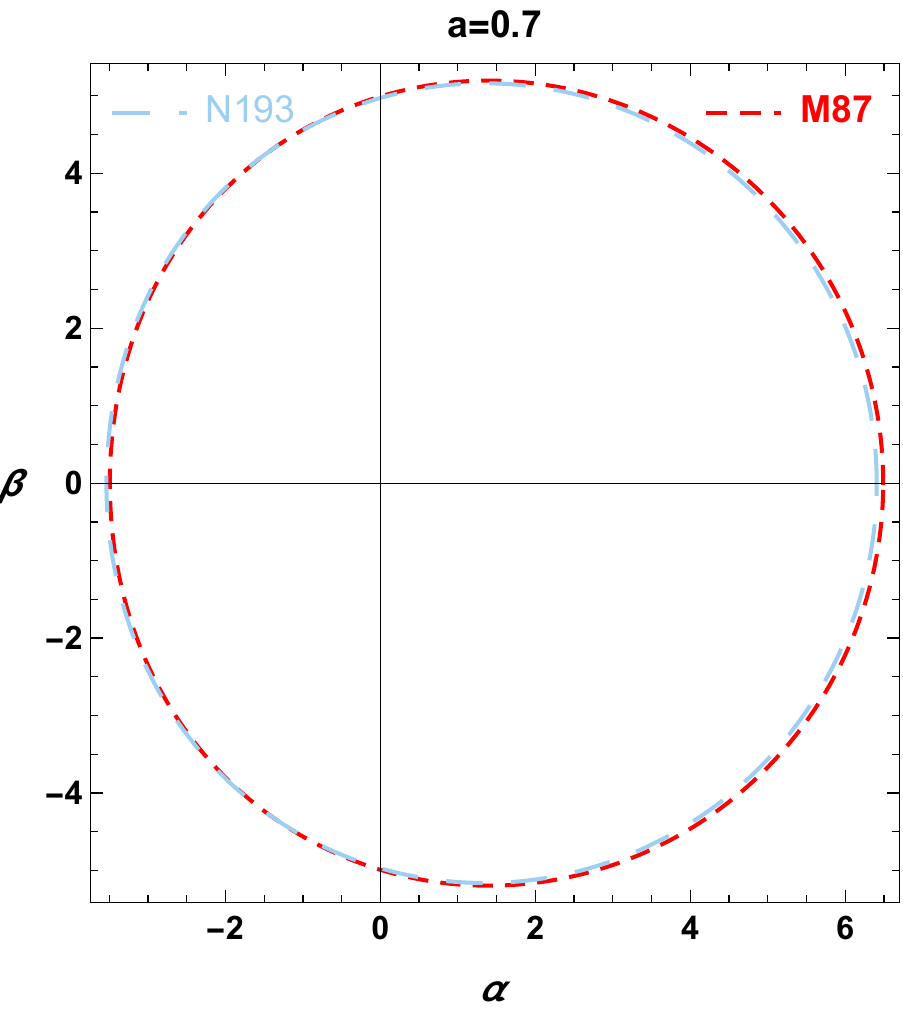} \>
			\includegraphics[scale=.47]{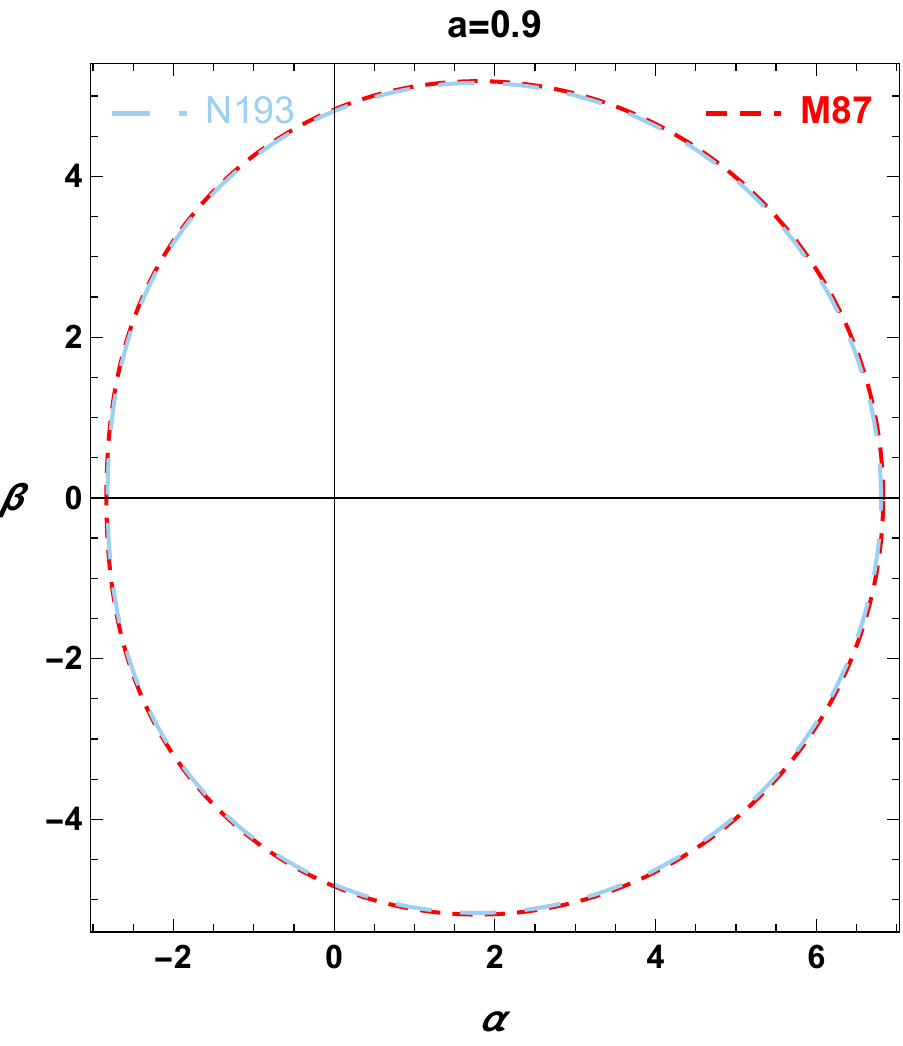} \\
				        \end{tabbing}
\caption{{\it \footnotesize  Black hole shadows for  $N=193$ compared by M87 shadow, for different values of $a$, using  M = 1 in units of the M87 black hole mass given by $M_{BH} =6.5\times 10^{9}M_\odot$ and $r_0 =91.2 kpc$}\cite{jusufi2019black}.}
\label{f2}
\end{figure}

\begin{figure}[h]
		\begin{center}
		\centering
			\begin{tabbing}
			\centering
			\hspace{8.4cm}\=\kill
			\includegraphics[scale=.38]{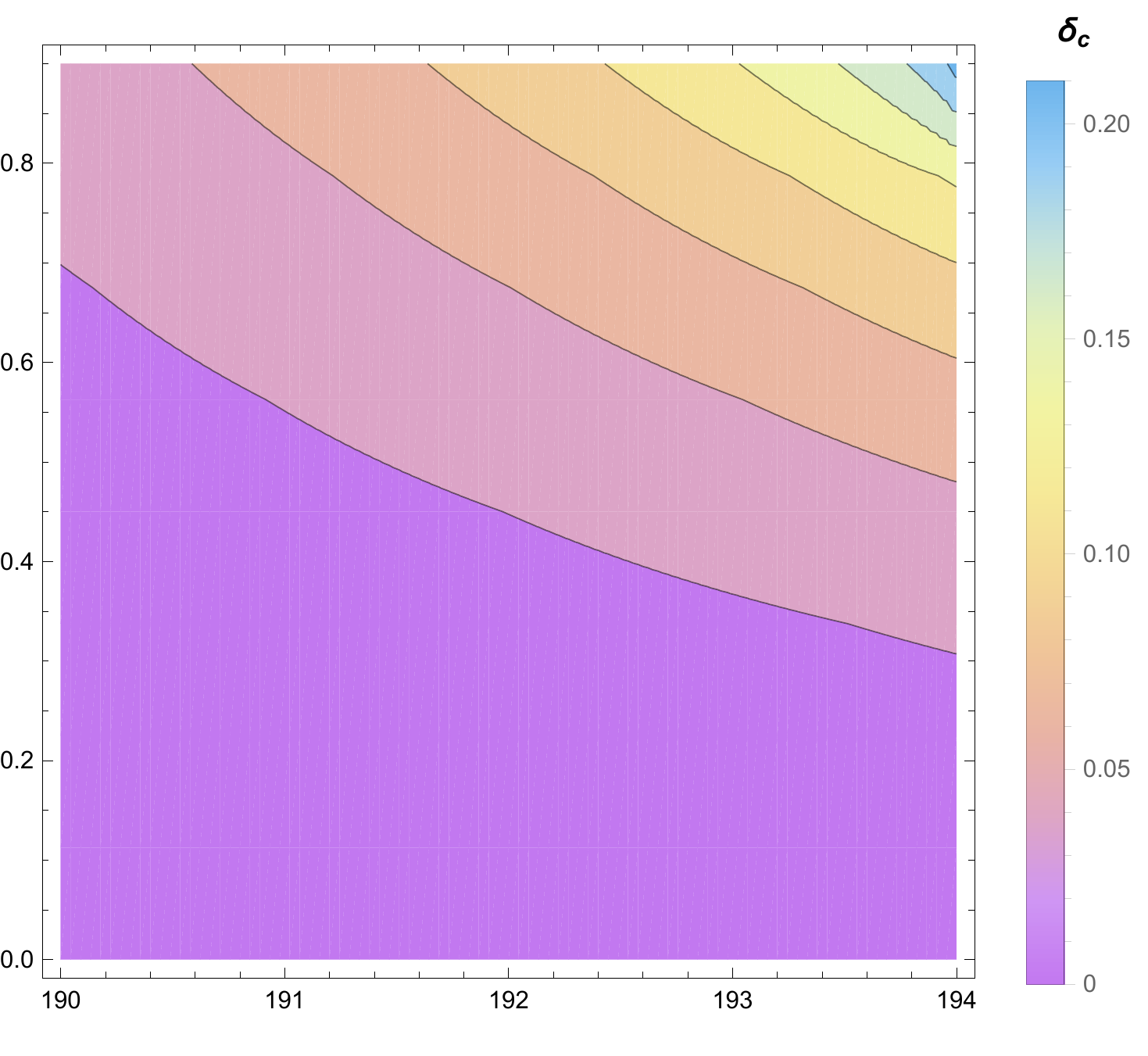} \>
			\includegraphics[scale=0.6]{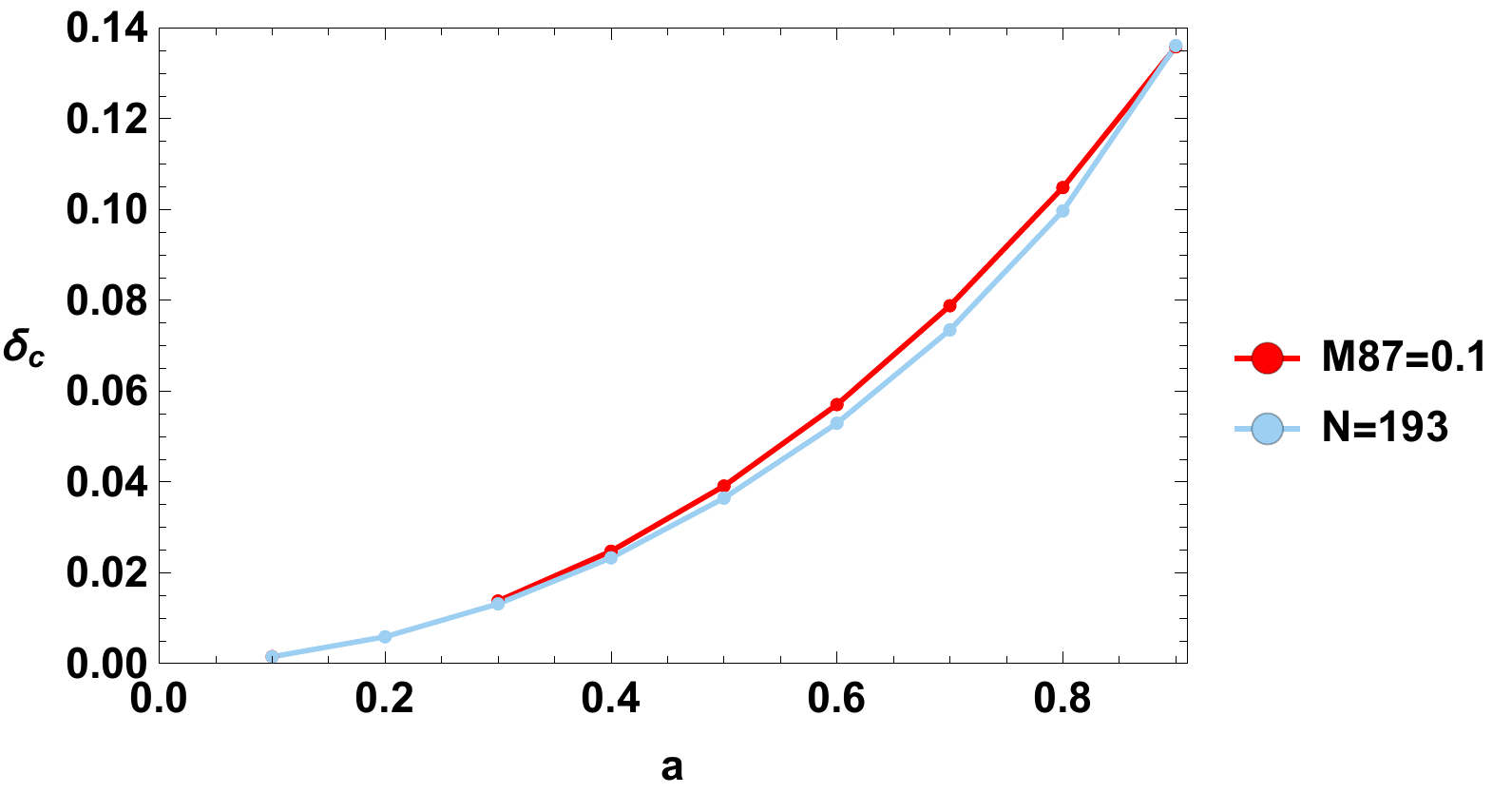}
		   \end{tabbing}
\caption{{
\it \footnotesize
Black hole distortion $\delta_c$ for  $N=193$ compared by M87 shadow, for different values of $a$, using  M = 1 in units of the M87 black hole mass given by $M_{BH} =6.5\times 10^{9}M_\odot$ and $r_0 =91.2 kpc$}.}
\label{diso1}
\end{center}
\end{figure}

Considering the  Fig.(\ref{f2}),  a close comparison of the radius of the both black holes  reveals that the  $M87^\star$ shadows could coincide perfectly with the M-theory one for  a  specific value of the brane number given by  $N=193$.  This could  constitute an observational constraint on M-theory  backgrounds. Knowing that the light of black hole  appears distorted,  it is crucial to use the distortion parameter to consolidate  such a  brane number  constraint.  In Fig.(\ref{diso1}),  we plot the distortion $\delta_c$ of  both black holes as a function of  the rotating  parameter $a$ and  the brane number $N$ around  the mentioned  specific value $N=193$ (left panel) and  the distortion $\delta_c$ as a function of $a$ and for $N=193$ (right panel).

 We notice from the left panel of Fig.(\ref{diso1}) that the distortion values for $190<N<194$ and for $a<0.6$ is almost null. However,  for rotation parameter value $a>0.6$,  the distortion increases by increasing   $a$ and $N$. In the right panel of Fig.(\ref{diso1}),  we observe that for the specific value of the  brane number $N=193$ and from the value of rotation parameter $a$,  the two distortion values for the black hole in M-theory and $M87^\star$ are almost the same. This  confirms the  specific  constraint in the used M-theory backgrounds.
\section{Discussions and concluding remarks}

In this work, we have  investigated the black hole shadows in M-theory inspired geometries,  for both rotating
and non-rotating solutions.
The present study has been made in terms of a reduced moduli space  coordinated  by the M2-brane number and the
spinning  parameter.     First,  we have   considered  the photon orbits  for non-rotating solutions by establishing
the associated shadow circular equations. In  particular, we have inspected the effect of the  M2-brane number.
Such effects have been  illustrated for a  particular situation relying on the equatorial plane. Concretely, we
have  realized that   the brane physics  could be exploited to control the size of such behaviors.
As expected, the rotation parameter   has been  used to  distort such a  circular black hole shape.
In certain regions of the moduli space,  we
have shown that  the shadows change from a D-shape to a  cardioid one.
A close examination   has  revealed  that
this heart shaped  could have some relations  with the cosmological constant and the  charge via
brane physics.
 This observation,    being  explored   and  supported by the recent work \cite{our}, is the object of further research to
 be presented elsewhere. 
 In addition,  we have promoted  the role of the spin  quantity from a  parameter
 to a variable. In such a  view,  the   shadows  have been  viewed as a two dimensional  object
 formed by   a  one dimensional geometry  fibered over  the interval describing  the rotating spin quantity.
  In this way,  the M2-brane number controls the size of the shadow fibers.
  Next, we have studied the energy emission rates in  certain geometric optical limits.
   Assuming that the area of  the black hole shadow is approximatively the high-energy
absorption cross section, we have  approached the energy emission rate.
The present results have  revealed  that  the M2-brane number effect is the same for the rotating and the non-rotating cases decreasing  the energy  emission rate.
It has been  shown that the M2-brane number could be considered as a tuning tool to deal with
 black hole stabilization issues. For the rotating case, some distinctions  appear. Increasing the rotation rate,
  the emission rate decreases. At the end, using the observational data associated with $M87^\star$,  we  have derived  an experimental constraint for which  the  $M87^\star$  shadow can be modeled with a rotating black hole in the M-theory background if the number of the M2-brane is exactly $N=193$.

This  work opens up for further studies. One interesting  question is to complete this analysis
by considering   external sources  involving dark sectors associated with axion stringy  fields \cite{boby17}. We hope to report elsewhere on  all mentioned  open questions.

\section*{Acknowledgment}
 AB would
like to thank the Departamento de F\'isica, Universidad de Murcia for very kind hospitality. The authors would like to thank  J. J. Fern\'andez-Melgarejo, Y. Hassouni, K. Masmar,  M. B. Sedra and A. Segui for discussions on related topics. They are
also grateful to the anonymous referee for their careful reading of our manuscript, insightful
comments, and suggestions, which have allowed us to improve this paper.
The work of ET  has been  supported in part by
the Spanish Ministerio de Universidades and Fundacion
Seneca (CARM Murcia) grants FIS2015-3454, PI2019-2356B and the Universidad de Murcia project E024-018. This work is partially supported by the ICTP through AF-13.

\end{document}